\documentclass[preprint]{aastex}
\usepackage{graphicx}
\usepackage{amsmath}

\shorttitle{Variable L dwarf DENIS 1058}
\shortauthors{Heinze et al.}

\begin{document}

\title{Weather on Other Worlds I: Detection of Periodic Variability in the L3 
Dwarf DENIS-P J1058.7-1548 with Precise Multi-Wavelength Photometry}

\author{Aren N. Heinze\altaffilmark{1}, Stanimir Metchev\altaffilmark{1}, Daniel Apai\altaffilmark{2,3}, Davin Flateau\altaffilmark{2,3}, Radostin Kurtev\altaffilmark{4}, Mark Marley\altaffilmark{5}, Jacqueline Radigan\altaffilmark{6}, Adam J. Burgasser\altaffilmark{7}, \'{E}tienne Artigau\altaffilmark{8}, and Peter Plavchan\altaffilmark{9}}

\altaffiltext{1}{Department of Physics and Astronomy, State University of New York, Stony Brook, NY 11794-3800, USA; aren.heinze@stonybrook.edu, stanimir.metchev@stonybrook.edu}
\altaffiltext{2}{University of Arizona Department of Astronomy, 933 N. Cherry Avenue, Tucson, AZ 85721, USA}
\altaffiltext{3}{University of Arizona Department of Planetary Sciences and Lunar and Planetary Laboratory, 1629 E University Blvd, Tucson, AZ, 85721, USA}
\altaffiltext{4}{Departamento de Fisica y Astronomia,
Facultad de Ciencias, Universidad de Valparaiso, Av. Gran Bretana 1111, Casilla 5030, Valparaiso, Chile}
\altaffiltext{5}{NASA Ames Research Center, MS-245-3, Moffett Field, CA 94035; Mark.S.Marley@NASA.gov}
\altaffiltext{6}{Department of Astronomy and Astrophysics, University of Toronto, 50 St. George Street, Toronto, ON M5S 3H4, Canada}
\altaffiltext{7}{University of California San Diego, Center for Astrophysics and Space Science, 9500 Gilman Drive, Mail Code 0424, La Jolla, CA 92093, USA}
\altaffiltext{8}{D\'{e}partement de Physique and Observatoire du Mont M\'{e}gantic, Universit\'{e} de Montr\'{e}al, C.P. 6128, Succ. Centre-Ville, Montr\'{e}al, QC, H3C 3J7, Canada}
\altaffiltext{9}{NASA Exoplanet Science Institute, California Institute of Technology, M/C 100-22, 770 South Wilson Avenue, Pasadena, CA 91125, USA}

\begin{abstract}
Photometric monitoring from warm Spitzer reveals that the L3 dwarf
DENIS-P J1058.7-1548 varies sinusoidally in brightness with a 
period of $4.25^{+0.26}_{-0.16}$ hours and an amplitude of $0.388 \pm 0.043$\% (peak-to-valley)
in the 3.6$\mu$m band, confirming the reality of a $4.31 \pm 0.31$ hour periodicity
detected in $J$-band photometry from the SOAR telescope. The $J$-band variations
are a factor of $2.17 \pm 0.35$ larger in amplitude than those at 3.6$\mu$m,
while 4.5$\mu$m Spitzer observations yield a 4.5$\mu$m/3.6$\mu$m amplitude
ratio of only $0.23 \pm 0.15$, consistent with zero 4.5$\mu$m variability.  
This wide range in amplitudes indicates rotationally modulated variability
due to magnetic phenomena and/or inhomogeneous cloud cover.
Weak H$\alpha$ emission indicates some magnetic activity,
but it is difficult to explain the observed amplitudes
by magnetic phenomena unless they are combined with cloud
inhomogeneities (which might have a magnetic cause).  
However, inhomogenous cloudcover alone can explain all our observations,
and our data align with theory in requiring that the regions with
the thickest clouds also have the lowest effective temperature.
Combined with published $v \sin(i)$ results, our rotation period yields a 95\% confidence lower
limit of $R_* \ge 0.111~ R_{\sun}$, suggesting upper limits of 320 Myr and
0.055 $M_{\sun}$ on the age and mass.  These limits should be
regarded cautiously because of $\sim 3\sigma$ inconsistencies with other data; however,
a lower limit of $45^{\circ}$ on the inclination is more secure.
DENIS-P J1058.7-1548 is only the first of nearly two dozen
low-amplitude variables discovered and analyzed by the Weather
on Other Worlds project.

\end{abstract}

\keywords{stars: individual (DENIS-P J1058.7-1548) --- stars: low-mass, brown dwarfs --- 
stars: rotation --- stars: spots --- stars: variables: other --- techniques: photometric}

\section{Introduction}

\subsection{L dwarfs, clouds, and the L/T Transition}

The L dwarfs \citep{Kirk} include brown dwarfs with ages from tens
of Myr to several Gyr (depending on mass) and also the coolest and
lowest mass main sequence stars \citep{Bur97, Bur06}.  
Together with the T and Y dwarfs and the latest M dwarfs, they make
up the ultra-cool dwarfs (UCDs), which have atmospheric temperatures low
enough for the formation of condensates.
In L dwarfs these take the form of thick clouds of refractory silicate
`dust' and liquid iron droplets, which profoundly influence the emitted spectra \citep{Allard01,ackerman}.
The presence of numerous condensing molecular species make modeling 
these objects a complex task \citep{Tsuji1,Allard97,Allard01,ackerman,Bur06,Helling08,marley,patchy01}, 
but L dwarf clouds also hold the potential for fascinating
phenomena such as molten iron `rain', hot silicate `snow' (see for example Ackerman \& Marley 2001),
and detectable weather patterns analogous to Jupiter's Great Red Spot \citep{GelinoThesis,SIMP0136}.
L dwarfs may also allow us to study young giant planets by analogy, given
their similar values of effective temperature ($T_{\mathrm{eff}}$).



The silicate and iron clouds of L dwarfs form a global overcast near or above the
photosphere.  They redden the emitted
spectrum \citep{Allard01,ackerman,Knapp04}, and
apparently produce the 9-11 $\mu$m absorption attributed to silicate
grains in Spitzer/IRS spectra of L dwarfs \citep{Cushing06,Burgasser08,Looper08}.  As objects become
cooler toward the L-T transition, the effects of the iron and silicate clouds
diminish, resulting in bluer near-infrared (near-IR) colors.  
This blueward shift happens over a very small range in $T_{\mathrm{eff}}$,
and over this range the $J$-band luminosity actually increases
with decreasing temperature (Dahn et al. 2002; Tinney et al. 2003; Vrba et al. 2004;
see also Knapp et al. 2004).  This cannot be explained
by the clouds simply sinking below the photosphere with decreasing $T_{\mathrm{eff}}$ \citep{FeH,Bur06}.  
The clouds must additionally break up \citep{ackerman,FeH}, and/or rain out \citep{ackerman} as
$T_{\mathrm{eff}}$ decreases.  This suggests that L/T transition objects
may have patchy clouds.



Patchy clouds on UCDs should cause rotationally
modulated photometric variability due to flux differences between the
most-cloudy and least-cloudy hemispheres.  Numerous searches for
variability in L and T dwarfs have been performed to date
\footnote{E.g. \citet{Bailer99,Bailer01,Kelu-1var,Clarke02,GelinoThesis,Gelino02,Clarke03,Enoch03,Koen03,KoenI,KoenJHK,Goldman05,Koen05,SpitzerPhot1,Lane07,Clarke08,Littlefair,SIMP0136,2M2139,Buenzli2012,Apai2013}.}.  The first
two objects found with large (5-30\%) amplitude periodic variability were indeed at the L/T transition: the 
T2.5 dwarf SIMP J013656.57+093347.3 \citep{SIMP0136,Apai2013} and the 
T1.5 dwarf 2MASS J21392676+0220226 \citep{2M2139,Apai2013}.

However, variability at lower amplitudes has been seen in L dwarfs
that, like DENIS-P J1058.7-1548 (hereafter DENIS 1058), are far from the L/T transition
\citep{Kelu-1var,Gelino02,KoenI,Lane07}.  Magnetic starspots are considered as
a possible cause of variations by \citet{Kelu-1var} and \citet{Lane07},
with the latter preferring this explanation.
Magnetic phenomena (albeit emission regions rather than dark starspots)
certainly are the cause of radio and H$\alpha$ 
variations detected in some late-M and L dwarfs \citep{Hallinan07,Berger08,Berger09}.
Magnetic effects could also cause the continuum variability observed
in DENIS 1058 and in previous studies of L dwarfs.  However,
uneven clouds on a less extreme scale than is seen in L/T transition objects would
also be consistent with all the data, especially if the inhomogeneities
took the form of variations in cloud thickness rather than complete clearings in the global overcast.
When L dwarf spectra are fit by models including clouds with thickness
parametrized by the sedimentation or `rain' parameter $f_{\mathrm{sed}}$,
introduced by \citet{ackerman} and used widely since \citep{FeH,Knapp04,marley},
it is found that different objects are best fit by different values of $f_{\mathrm{sed}}$
\citep{Burgasser08,stephens}.  Similarly, there is no reason to assume that the global 
clouds of a given L dwarf will be specified
by a spatially and temporally unvarying value of $f_{\mathrm{sed}}$.


The photometric effects of inhomogeneous clouds in L and T dwarfs depend strongly
on the wavelength of observation.  This is due to the strong molecular gas
absorption in the atmospheres (from H$_2$O and CH$_4$), which causes 
the effective altitude (and thus temperature) of the photosphere to vary
from one wavelength to another \citep{Bur97,ackerman,LT,Bur06}.  
This wavelength dependence makes multi-band photometric monitoring a useful
tool for understanding the vertical structure of clouds in UCDs.

\subsection{The L3 Dwarf DENIS-P J1058.7-1548} \label{sec:targ}

Herein we report observations of DENIS 1058 made
as part of the Weather on Other Worlds (WoW) project, a Spitzer Exploration Science
program probing the photometric variability of brown dwarfs (Metchev et al. 2013, in prep).
DENIS 1058 is the first WoW target for which we acquired ground based photometry
nearly simultaneous with the Spitzer observations.

DENIS 1058 \citep{discovery} is an L3 dwarf \citep{Kirk} at a distance of
$17.33 \pm 0.30$ pc \citep{Dahn02}.  At $J=14.155 \pm 0.035$ mag and $J-K_S=1.623 \pm 0.045$ mag \citep{2MASS}, its
brightness and colors are normal for its spectral type. High resolution 
HST/NICMOS images with good sensitivity down to separations of 0.1 arcsec
show no evidence of a binary companion \citep{reid08}. 
Near-IR spectroscopy and $L'$-band photometry indicate
a bolometric flux of $1.10 \times 10^{-14}$ W m$^{-2}$ \citep{spec01}, with
\citet{Dahn02} finding a consistent value.

\citet{martin99} give $T_{\mathrm{eff}} = 1900$ K based on optical spectroscopy,
while \citet{Teff} find $T_{\mathrm{eff}} = 1950$ K.  \citet{Dahn02}
use a bolometric luminosity derived from the measured $K_S$-band flux to arrive
at a consistent result of $T_{\mathrm{eff}} = 1945 \pm 65$ K (based on a radius of $0.0903 R_{\sun}$, 
obtained by averaging theoretical radii corresponding to the bolometric luminosity at ages of 1.0
and 5.0 Gyr).  DENIS 1058 has an age of at least a few hundred Myr based on non-detections of the
6708 \AA~ lithium line down to upper limits in the equivalent width of 
0.3--0.5 \AA~ \citep{noLi,martin97,Kirk}.  \citet{martin97} comment that DENIS 1058's spectral
characteristics could be consistent with a very low mass
($0.075 M_{\sun}$) star of age 3 Gyr or an $0.065 M_{\sun}$ brown dwarf of age 800 Myr.
They prefer the latter possibility based on their radial velocity measurement
of $11 \pm 5$ km s$^{-1}$, which, like the tangential velocity of $21.1 \pm 0.4$ km s$^{-1}$
measured later by \citet{Dahn02}, is relatively low and therefore consistent with youth.

Of particular relevance to the measurements we report herein of DENIS 1058's
photometric rotation period, \citet{martin97} measure its projected rotational
velocity from line-broadening in their Keck HIRES spectrum, obtaining
$v \sin(i) = 30.0 \pm 10.0$ km s$^{-1}$.  \citet{Teff} perform a more sophisticated
analysis of the same spectrum to find $v \sin(i) = 37.5 \pm 2.5$ km s$^{-1}$.

Denis 1058 exhibits weak H$\alpha$ emission \citep{noLi,Kirk,martin99},
which indicates some magnetic activity.  However, the ratio of H$\alpha$
to bolometric luminosity is extremely low, with $\log(\frac{L_{H\alpha}}{L_{bol}}) = -5.67$ \citep{flare}.
Thus DENIS 1058 is not an exception to the trend of magnetic activity diminishing
greatly with decreasing $T_{\mathrm{eff}}$, which has
been explored by \citet{flare} and others.  While H$\alpha$ emission in UCDs 
can accompany radio emission (e.g., Berger et al. 2009), we are not aware of any 
published radio observations of DENIS 1058.

\section{Observations and Data Reduction} \label{sec:Data}
Our data consist of 20 hours of photometric monitoring of DENIS 1058: first 7 hours in the $J$-band at
the 4m SOAR telescope and then, three
days later, 7 hours in the Spitzer IRAC 3.6$\mu$m bandpass (hereafter [3.6]) followed
immediately by 6 hours in the IRAC 4.5$\mu$m bandpass (hereafter [4.5]).  Observation specifics
are presented in Table \ref{tab:data}.  

\begin{deluxetable}{ccccccc}
\tablewidth{0pc}
\tablecolumns{7}
\tablecaption{DENIS 1058 photometric monitoring data acquired \label{tab:data}}
\tablehead{ & \multicolumn{2}{c}{Beginning of Observations} & \colhead{Duration of}  & \colhead{Exposure} & \colhead{Images} & \colhead{Per-image} \\
\colhead{Wavelength} & \colhead{HMJD\tablenotemark{a}} & \colhead{UTC Date \& Time} & \colhead{Monitoring} & \colhead{Times} & \colhead{Acquired}& \colhead{RMS error\tablenotemark{b}} }
\startdata
$J$-band (1.25$\mu$) & 56000.030 & 2012-03-14, 00:36 & 7.01 hr & 100 s & 159 & 0.30\% \\
IRAC [3.6] & 56003.036 & 2012-03-17, 00:47 & 7.33 hr & 12 s\tablenotemark{c} & 2166 & 0.59\% \\
IRAC [4.5] & 56003.367 & 2012-03-17, 08:44 & 5.85 hr & 12 s\tablenotemark{c} & 1619 & 0.76\% \\
\enddata
\tablenotetext{a}{\footnotesize{HMJD = Heliocentric Modified Julian Day}}
\tablenotetext{b}{\footnotesize{Note that if the exposure
time difference is taken into account, both IRAC bands offer precision 
superior to that at $J$; ground-based $J$-band measurements typically also suffer
from greater systematic error.}}
\tablenotetext{c}{\footnotesize{This is the time spent integrating
the whole array and is approximately equal to the sampling time.
The per-pixel integration time is slightly shorter at 10.4 seconds}}
\end{deluxetable}

IR detectors on both Spitzer and ground based telescopes are known
to suffer from both inter-pixel and intra-pixel sensitivity variations
which can create systematic errors in time-series photometry.  As outlined
below, our data
acquisition strategies were designed to minimize such errors.

\subsection{Spitzer Data Acquisition} \label{sec:spdata}

\begin{figure}
\includegraphics[scale=0.6]{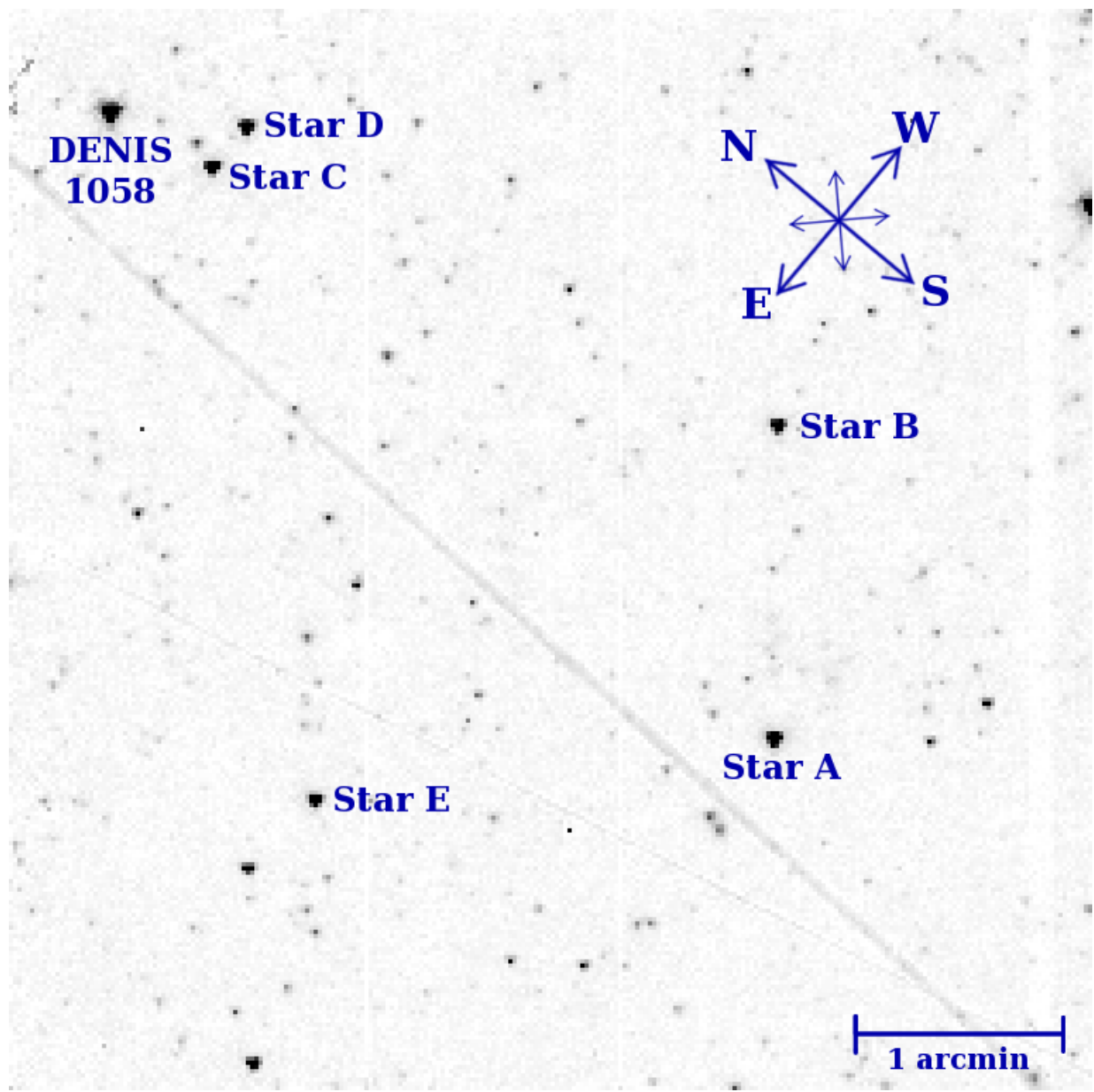}
\caption{Stacked image made from all frames of our IRAC [3.6]
data.  Stars used to measure systematic errors are labeled.
The streaks crossing the frame are not asteroids, because their
appearance on each individual frames matches that on the full stack.  
They may be residual images of bright stars across which Spitzer
has panned in its acquisition slew.  The appearance of the starfield in the [4.5] image is similar.
\label{fig:IRACfield}}
\end{figure}

For our observations with Spitzer IRAC \citep{IRACref}, we used the `staring mode' that has become
standard for precision time-series photometry \citep{staring1}, in which no dithering
or other intentional alteration of the telescope pointing is carried out
during the observing sequence.  Although we were using full-frame images
from IRAC, we elected to position DENIS 1058 in the upper left
corner of the detector, which is the region used for subarray observations
of bright targets.  This was intended to give us the option of using the
`sweet spot' calibration from the Spitzer Science Center -- that is, a
carefully measured sensitivity map of a specific pixel in the subarray
recommended for use with precision photometry.  The `sweet spot'
map is still under development, and calibrations based on it have not
benefited our photometric precision thus far; the methods we
have actually used to remove the effects of intra-pixel sensitivity
variations are described in \S~\ref{sec:ch1s}.  The per-image random
errors in our photometry of DENIS 1058 are 0.59\% and 0.76\% for the
IRAC [3.6] and [4.5] bands, respectively, calculated from the
differences between adjacent photometric points (a difference which
measures pure random noise because no known astrophysical
or systematic effects are able to change the flux appreciably over the
12-second sampling interval).  These measured values compare well
with the calculated photonic shot-noise of 0.54\% and 0.63\%
respectively: the agreement indicates that read noise, randomly varying
aperture losses, and other effects make only
minor contributions to the photometric error.
Figure \ref{fig:IRACfield} shows a co-added image of all our [3.6] exposures.


\subsection{SOAR $J$-band data acquisition} \label{sec:Jdata}

\begin{figure}
\includegraphics[scale=0.45]{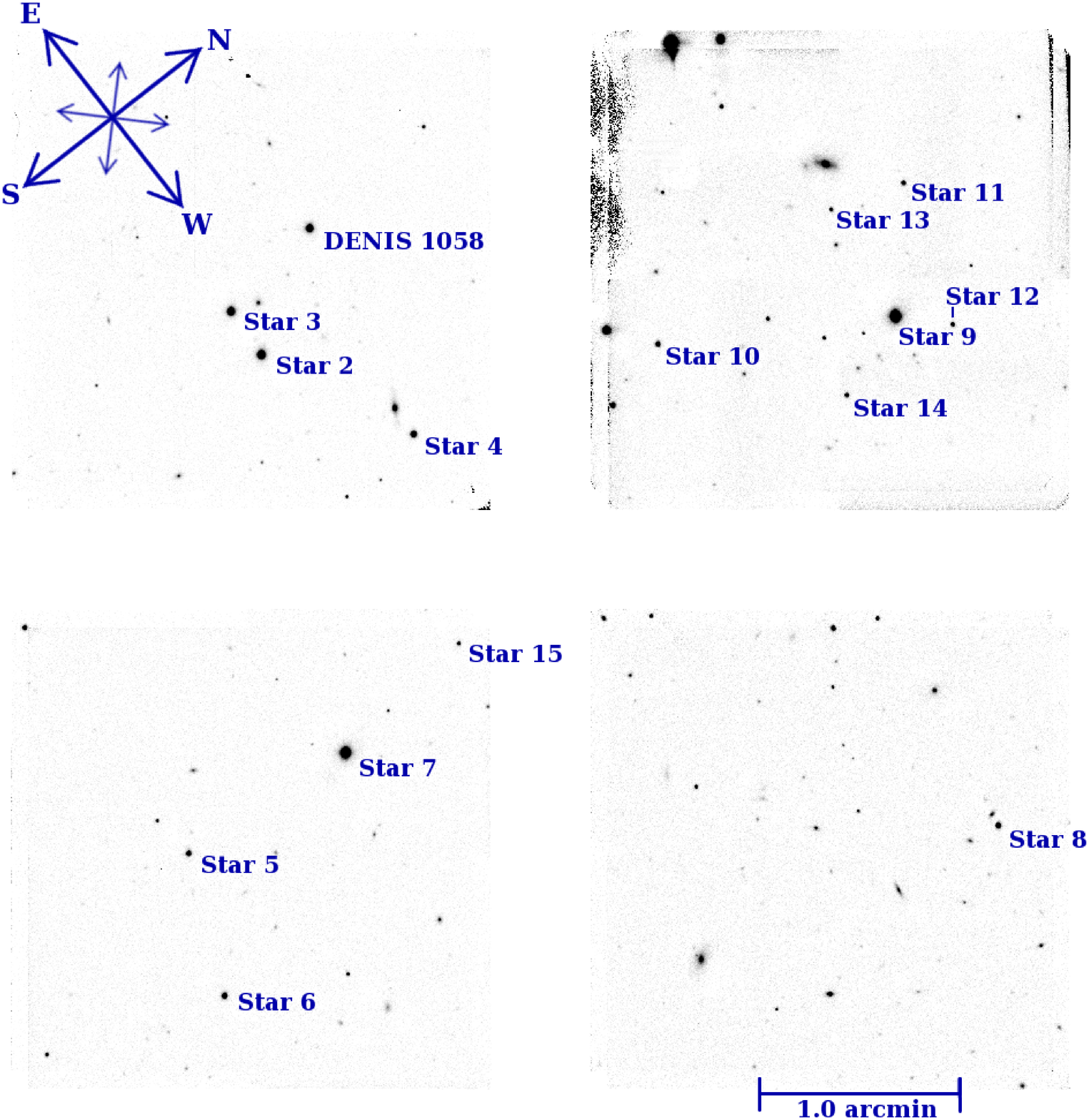}
\caption{Stacked set of 146 of our $J$-band images of the
DENIS 1058 field, with stars we considered as possible
photometric references labeled.  Note the separate images
from the four detectors of the Spartan IR Camera, with
substantial gaps that we did not attempt to fill by
dithered imaging.  The object immediately to the upper left of 
Star 4 is a galaxy; its proximity may have slightly
affected the Star 4 photometry in one of the nod positions,
but we believe we have successfully modeled and corrected
for this effect.  Note that Stars 2 and 3 here are Stars D and C,
respectively, in Figure \ref{fig:IRACfield}.  Other than this
there is no overlap between the sets of comparison stars.
Noisy artifacts at the vertical edges of the upper right detector
are due to insensitive regions on the chip, which have no
effect on any measured star.
\label{fig:Jfield}}
\end{figure}

For our $J$-band observations with the SOAR telescope, we used the Spartan IR
Camera \citep{SpartanCam} in its wide field configuration.  The Spartan
instrument uses four 2048x2048 HAWAII-2 detectors to deliver a total field
of view of about 5x5 arcmin at a pixel scale of 0.068 arcsec/pixel, with
roughly 0.5 arcmin gaps between the detectors.  Figure \ref{fig:Jfield} shows
a coadded image of our Spartan data.  At the beginning of our observations the background level
was changing rather quickly, so we elected to nod the telescope by 7 arcsec
every 2 images, rather than take most of the images at a single pointing and
interleave occasional dithered sequences.  The exquisite pointing
and tracking precision of the SOAR telescope allowed us to keep the two nod
positions consistent with an RMS pointing variation of only about one pixel
(0.07 arcsec). In a bid for potentially better photometric precision, we 
also took 14 of our 159 photometric images with the telescope defocused enough
to increase the effective FWHM of our images by a bit less than a factor of two.
However, as the mean FWHM of in-focus images was $\sim 10$ pixels
(0.7 arcsec), the images were already highly oversampled.  The defocusing
did not help; indeed, by necessitating a larger photometric aperture, it increased
the background noise and rendered the photometry inferior to that obtained
from in-focus images (though photometry from the defocused images remained usable).


Our images were processed by dark subtraction and flatfielding using
twilight sky flats (dome flats were tried, but proved much less effective
at removing the effects of dust shadows on the detector).  Cosmic rays hits
were numerous due to the long, 100 second exposures.  We removed
them using the laplacian edge detection algorithm of
\citet{lacos}, just after the flatfielding step.

Following cosmic ray removal, we performed sky subtraction on
each science image using another image taken close in time and in the opposite
nod position.  A scaling factor near unity was applied to each
nod-subtraction image to yield a zero-mean background for the subtracted
science image.  The final step in our processing was to merge the images
from the four detectors into a single master frame for each exposure; the
stack of all in-focus merged frames is shown in Figure \ref{fig:Jfield}.  Astrometrically,
the digital gaps between the images from different detectors are only
approximately correct.  The per-image error on our $J$-band photometry
of DENIS 1058 is 0.30\% (based on the RMS residuals from our final fit). This
is just over twice the per-image photonic shot-noise of 0.14\%, which
is consistent with our conclusion in
\S~\ref{sec:$J$-band} that unmodeled systematic effects remain in the $J$-band data.  As described
below, such effects are folded into our final uncertainties for the
$J$-band analysis.

\section{Spitzer IRAC [3.5] Data Analysis} \label{sec:analysis}

Confirming and characterizing astrophysical variability at an amplitude 
comparable to that of the systematic errors requires considerable analysis, which we detail
in this section.  Readers interested only in the final result should skip
to \S~\ref{sec:ch1m}.

In \S~\ref{sec:ch1p}, we describe our photometric methodology for IRAC images.  
In \S~\ref{sec:ch1a} we confirm the presence of astrophysical variability in
DENIS 1058 by first modeling and correcting for systematic effects and then performing a
Lomb-Scargle periodogram analysis, which outputs the false alarm probability (FAP),
a measure of the likelihood that any apparently coherent variations are due to
random noise.  Comparing the FAP of DENIS 1058 to that of identically
processed field stars demonstrates that DENIS 1058 is a variable.
However, the periodogram is not the best tool for a detailed analysis of
DENIS 1058's variations.
Instead, in \S~\ref{sec:ch1s} we fit them with a Fourier model using singular value
decomposition (SVD), test the robustness of the fit, and find a good
parameterization for the final 
Markov Chain Monte Carlo (MCMC) analysis needed to calculate uncertainties,
which is described in \S~\ref{sec:ch1m}.

\subsection{Photometry} \label{sec:ch1p}

\begin{figure}
\plotone{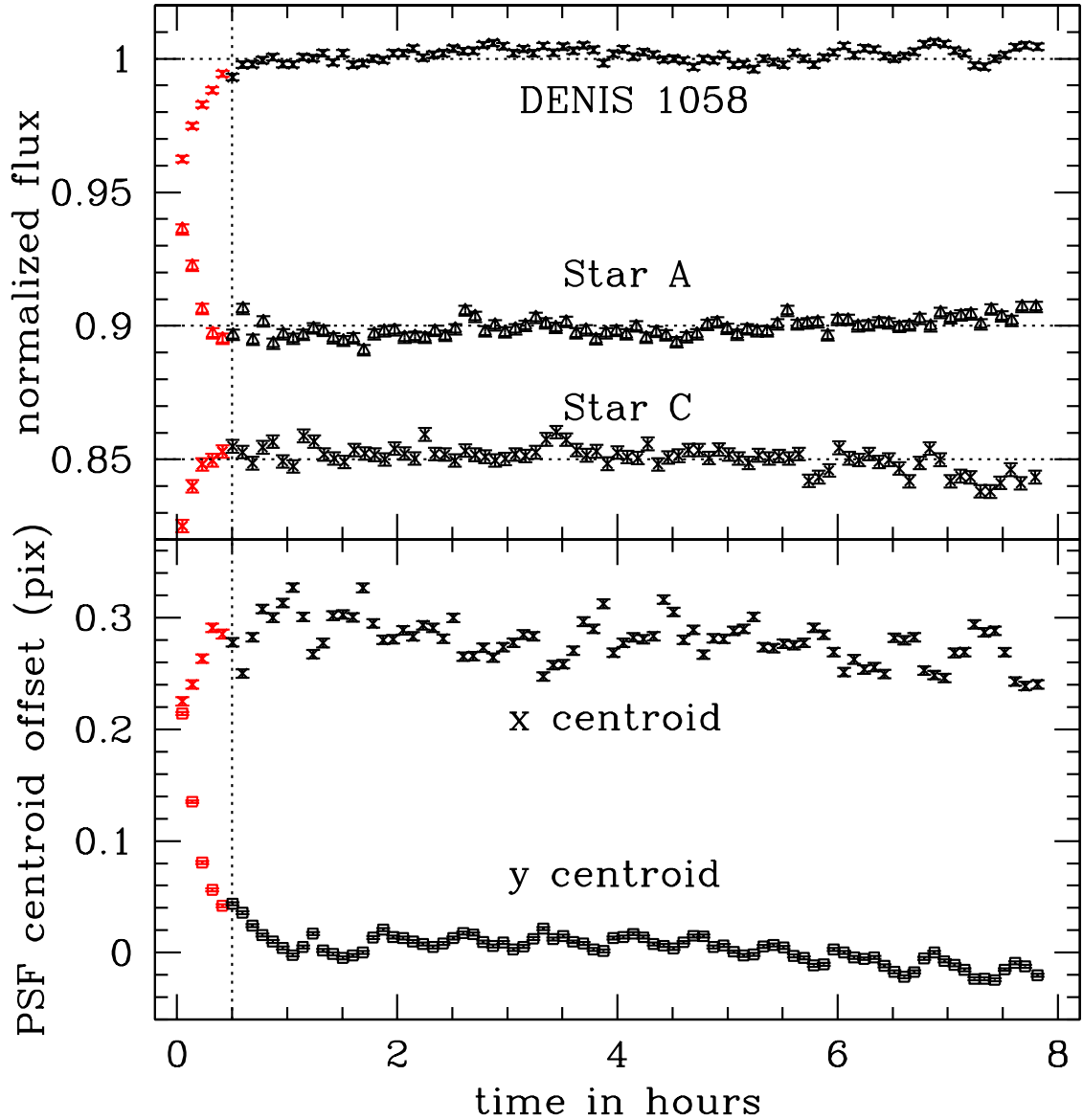}
\caption[IRAC ch1: raw photometry and image positions]{\textit{Upper Panel}:
Normalized uncorrected IRAC [3.6] photometry of DENIS 1058 and two comparison
stars.  For clarity, the data are binned in sets of 25 points (each bin
thus represents a 5-minute time interval) and the photometry for Star A and Star C
is offset by -0.1 and -0.15, respectively, relative to DENIS 1058.  
\textit{Lower Panel}: Location of the x and y centroids of DENIS 1058 
in the images.  An offset of 0,0 corresponds to the center of pixel 23,231 on the
IRAC [3.6] detector.  The centroid stayed on this pixel throughout the data
sequence, but its intra-pixel motion introduced systematic errors
in the relative photometry due to the pixel phase effect.  Previous to 
$t = 0.5$ hours (indicated by the vertical dotted line), the spacecraft
was still settling on its new pointing, and the data were not used in our
final fits.  Error bars are shown but in most cases are smaller than the symbols.}
\label{fig:ch1raw}
\end{figure}

Using IDL, we obtain photometry from Basic Calibrated Data images, provided
by the Spitzer Science Center after processing through IRAC pipeline version
19.1.0.  We obtain centroids by gaussian fitting with the \texttt{gcntrd} 
function, setting the FWHM to 4.5 pixels to minimize the scatter in stellar
positions. The fact that stellar images are actually much sharper than
4.5 pixels does not invalidate this choice because the \texttt{gcntrd} algorithm
uses it only to set the size of the fitting box.  We perform aperture
photometry about the measured centroids with an aperture radius chosen 
to minimize the RMS scatter of normalized photometry (2.1 pixels for DENIS 1058).  
Note that aperture photometry rather than point spread
function (PSF) fitting is normally used for IRAC images (e.g. \citet{SpitzerPhot1,Todorov12,np}); 
one reason for this may be IRAC's somewhat distorted `triangular' PSF.

To clip our photometric data, we first median-smooth each data vector with a sliding boxcar of width
25 points (6 minutes of time), and subtract this smoothed vector
from the original data.  All known relevant astrophysical or systematic
variations have characteristic timescales longer than six minutes,
so such signals should be absent in the
subtracted vector: thus, it can be screened for outliers without
danger of removing points at the extrema of variations we
wish to measure.  We identify outliers in the subtracted vector using the
\texttt{robust\_sigma} routine's default criterion for bad data, which (although based
on the median absolute deviation) corresponds approximately to a conservative,
6$\sigma$ clip.  Bad points are rejected
from the original data vector, and the surviving points (both photometry and
centroid positions) are binned in 10-point bins, resulting in a sampling
interval of about 120 seconds.  Figure \ref{fig:ch1raw} shows the
IRAC [3.6] photometry at this stage, combined with the pixel centroids.

The photometry shows the well known `pixel phase effect' in IRAC \citep{reach}: the measured flux
from an object depends on the object's exact position within a pixel.  
The anomalous photometry near the beginning of the observing sequence is
due to the settling of the telescope
pointing after its acquisition slew, and caused us to
reject the first 0.5 hours of [3.6] data
from our analysis.  We note that only the pixel phase effect seems
to be involved in producing this initial anomaly.
It cannot be due to the asymptotic
ramp phenomenon that has been reported in IRAC photometry of transiting
planetary systems (see for example \citet{Knutson09} and \citet{Todorov12}),
because the effect we observe does not have the same sign for all objects.
Similarly, we have not detected any effect corresponding to the linear time trends 
distinct from the pixel phase effect which have previously
been seen \citep{Deming12, Todorov12}.  The absence of such effects in our
data may be connected to the fact that our WoW targets are considerably
fainter than the majority of transiting planetary systems targeted by Spitzer.


 Following the initial settling, Spitzer's pointing shows a slow
linear drift and an oscillation with a period of 0.7-1.0 hours.  
Both effects are reflected in the photometry, which makes correcting the
pixel phase effect essential even after trimming the first 0.5 hours.

\subsection{Testing for Astrophysical Variability} \label{sec:ch1a}

The pixel phase effect can in principle be covariant with 
astrophysical variability.  Once we have established the presence of 
such varibility, we address this complication by performing a 
simultaneous pixel phase and variability fitting to the light curve 
(\S~3.3).  However, without a priori knowledge of the presence of 
astrophysical variability, we first implement a model only of the 
pixel phase effect and ratio it into our photometry.

\begin{figure}
\plotone{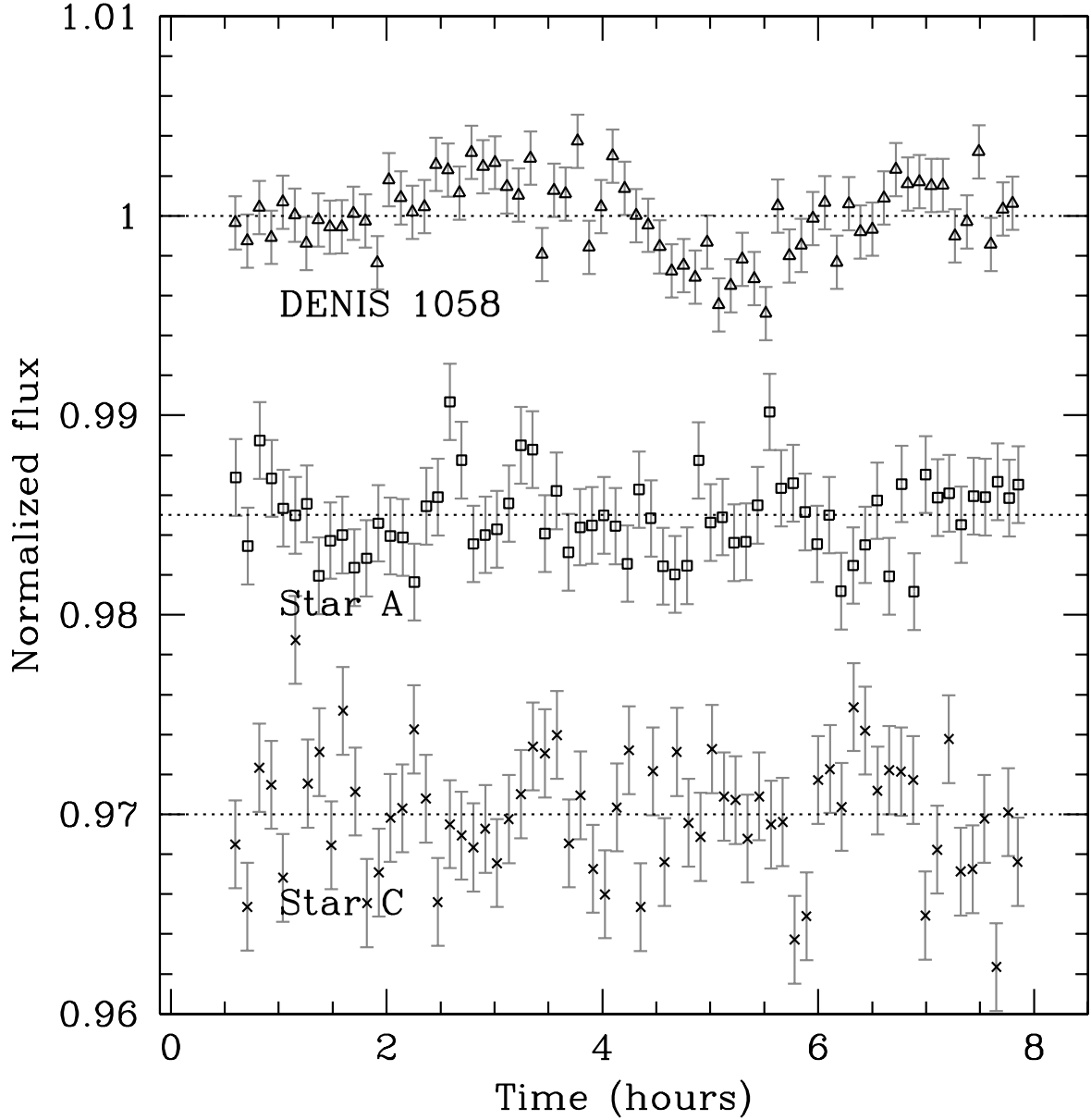}
\caption[IRAC ch1: Photometry after the `pixel phase' correction]{Normalized IRAC [3.6]
photometry of DENIS 1058 and the two brightest stars in the same IRAC field, after trimming
the first 30 minutes and correcting for the pixel phase effect using 
Equation \ref{eq:ppc01}.  Note that this figure has a much finer
vertical scale than Figure \ref{fig:ch1raw}. For clarity, the points here
are 30-point (6-minute) bins relative to the raw data, and
photometry for Stars A and C is offset by -0.015 and -0.03, respectively.
Linear slopes and short-period oscillations present in the uncorrected data have
vanished and a longer period, roughly sinusoidal variation emerges for DENIS 1058.
The error bars are based on the single-point photometric errors, scaled down
by $\sqrt 30$ due to the binning, with an additional (minor) contribution
due to the uncertainty on the centroids used in the pixel phase correction.}
\label{fig:ppc01}
\end{figure}

We model the pixel phase effect as a function of both the $x$ and $y$
pixel positions (see for example \citet{quadppc}).  Having tested functions linear
in $x$ and $y$ and found them insufficient, we choose to fit 
a quadratic function of the form:

\begin{equation}
f(x,y) = A_0 + A_1x + A_2y + A_3xy+ A_4x^2 +A_5y^2,
\label{eq:ppc01}
\end{equation}

where $f(x,y)$ models the measured flux, the $A_i$ are the fit coefficients, 
and $x$ and $y$ are sub-pixel coordinates (that is, the coordinates of the
object's centroid on a given image minus the coordinates of the pixel-center
nearest the object's average position over all the images). 
Although they move in sub-pixel coordinates, the centroids of
most objects are found on the same pixel for all images, so 
$x$ and $y$ are normally confined to the range $(-0.5,0.5)$.  We have confirmed
that our fit remains effective even when this is not the case, based on results
from stars in the field of DENIS 1058 and many other WoW targets.
We correct our photometry by dividing by $f(x,y)$;
Figure \ref{fig:ppc01} shows the resulting photometry for DENIS 1058 and
the two brightest field stars.  
DENIS 1058 exhibits roughly sinusoidal
variability, while the photometry of the stars shows little evidence
of coherent variations. We note that the uncertainties in the
measured $x$ and $y$ centroids of DENIS 1058 (per-image values about 0.019 and 0.007 pixels,
respectively) make only a minor contribution to the errors in the
corrected photometry, and the same is true of the field stars.

\begin{figure}
\plotone{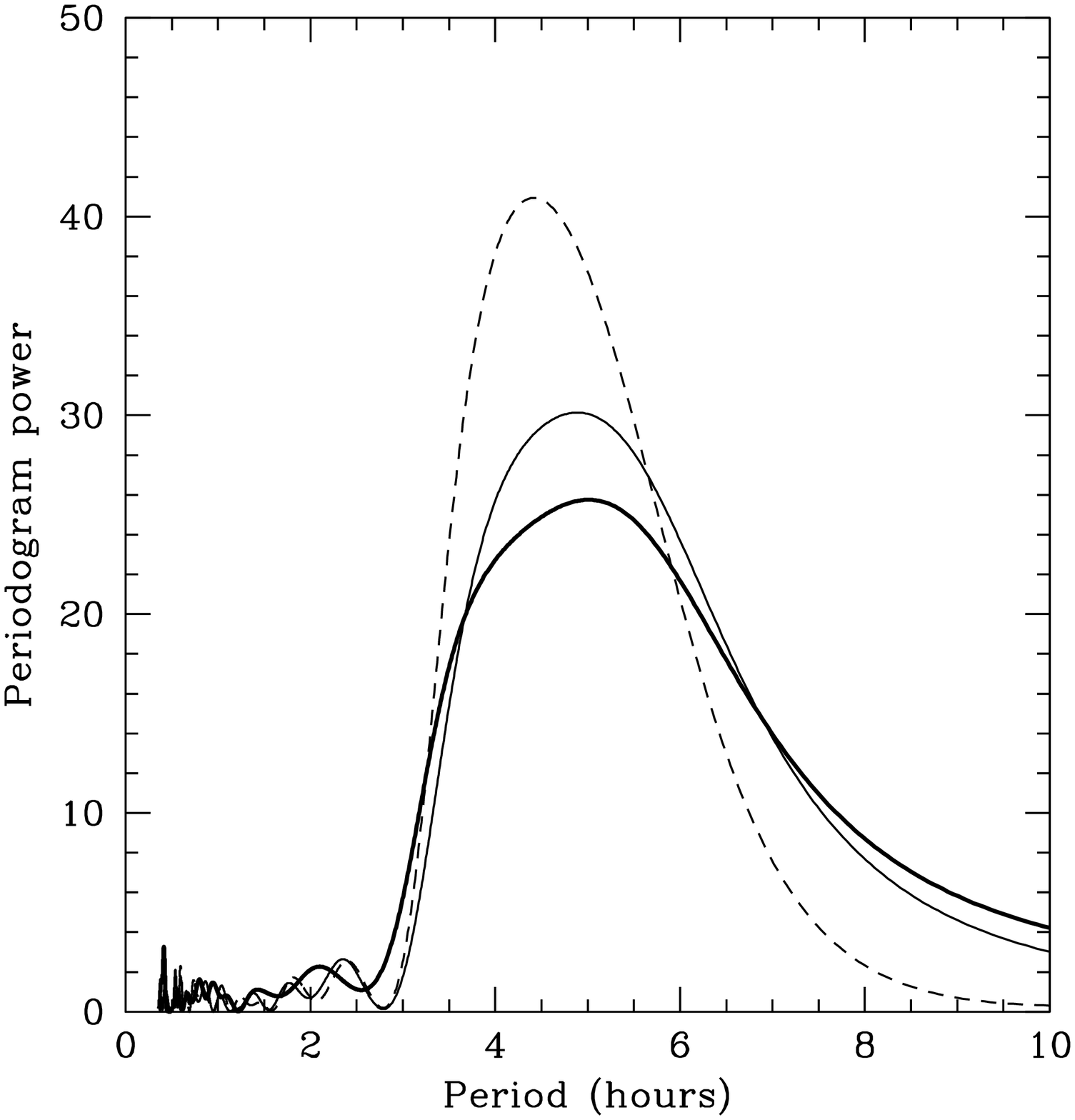}
\caption{Periodograms of real and simulated IRAC [3.6] photometry of DENIS 1058.
Heavy continuous line: periodogram of corrected real data.  Dashed line: periodogram
of simulated data consisting of the best-fit perfect sinusoid found in 
\S ~\ref{sec:ch1s}, with gaussian errors added having standard deviation
equal to the measured RMS.  Light continuous line: periodogram of simulated
data after `correction' for the pixel phase effect based on the actual 
measured centroids of DENIS 1058.  The pixel phase correction has distorted
the synthetic data such that the periodogram yields an inaccurate, 
longer period, and it appears that
the same thing has happened to the real data.  The fitting methods applied
in \S~\ref{sec:ch1s} and \S~\ref{sec:ch1m} are not subject to this bias.
Note that the width of the periodogram peaks is not
trivially related to the uncertainty of the period determination.
}
\label{fig:pgram}
\end{figure}

We probe the significance of DENIS 1058's apparent variability
by subjecting the corrected data to a Lomb-Scargle periodogram analysis,
as implemented by \citet{nrc}.  For our densely
and evenly sampled data, we oversample the periodogram by a factor of 200,
but probe only up to frequencies 5 times lower than the Nyquist.  As our analysis uses
binned data with 120-second sampling, the highest frequency we probe corresponds
to a 20 minute period.  The periodogram of our IRAC [3.6] photometry of DENIS
1058 is the heavy line in Figure \ref{fig:pgram}.

\begin{figure}
\plotone{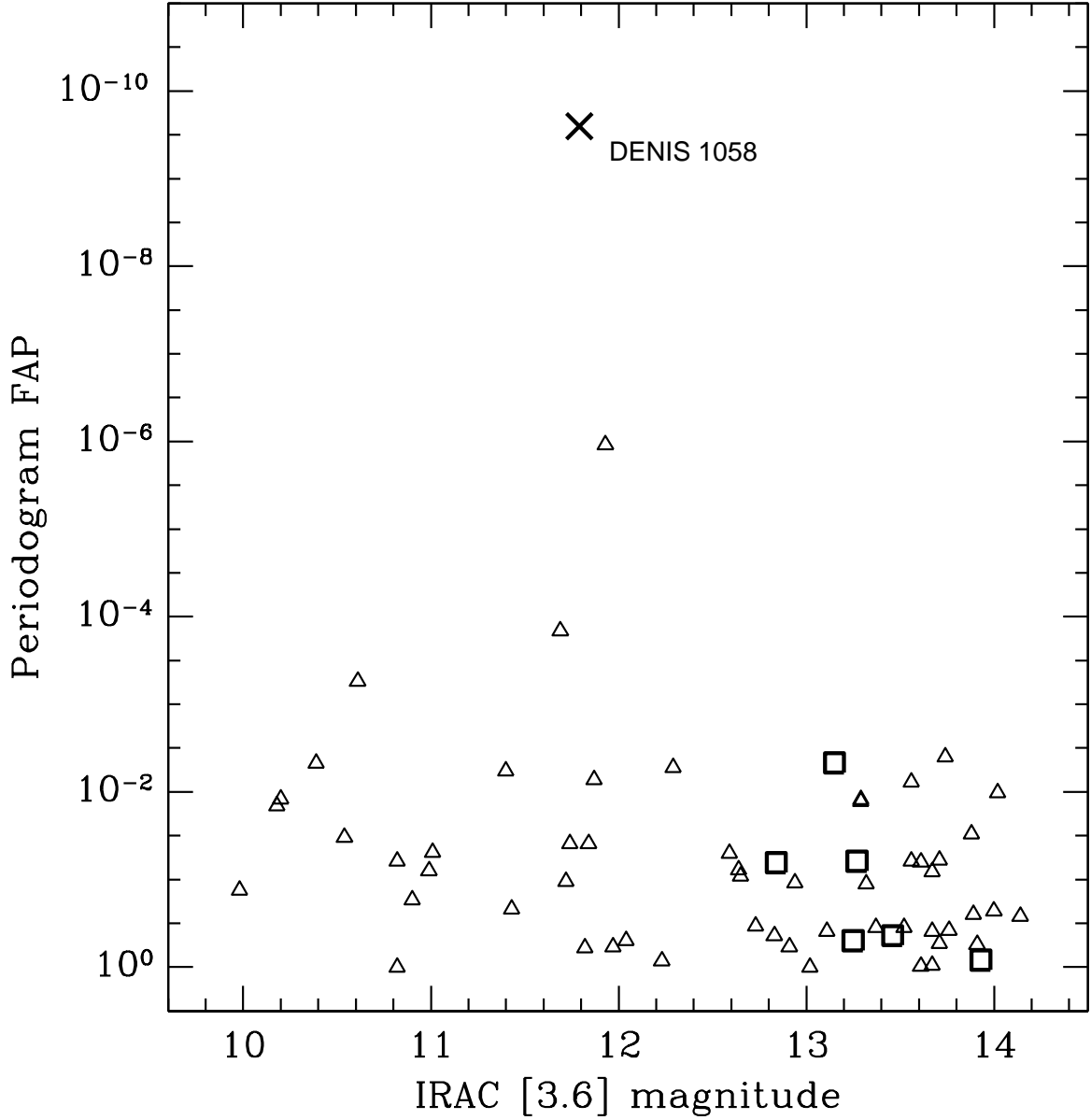}
\caption{Periodogram FAP vs. IRAC [3.6] magnitude
for field stars measured in WoW data (open symbols), and for DENIS 1058 (large `X').  
DENIS 1058 shows much more significant variations than any of 61 similar-brightness 
stars measured in 8 different WoW target fields.  Stars in the same field
as DENIS 1058 are shown as squares; the distribution of their FAP values
is consistent with that of the stars in the other fields.
}
\label{fig:FAP}
\end{figure}

Figure \ref{fig:FAP} shows the FAP values of DENIS 1058 and 61 comparable-brightness,
identically-processed stars from eight different WoW fields.  Although weak
residual systematics prevent the FAP values from being formally accurate (i.e.,
a periodogram FAP of $5 \times 10^{-2}$ does not imply a 95\% confidence
detection of variability), the figure shows that fewer than one in
fifty stars has a FAP value below $10^{-4}$; thus any object that does is a
genuine variable with a confidence level of $\sim$ 98\%.  
We note that no suspected variables among the field stars have been removed
from Figure \ref{fig:FAP} (only one previously-known eclipsing binary), so weak astrophysical
variations rather than residual systematics
could be responsible for the most significant apparent detections
among the stars.

With a FAP of $2.63 \times 10^{-10}$, the detection of astrophysical variability
in DENIS 1058 is unambiguous.  Table \ref{tab:pgram01} gives the FAP values
and IRAC magnitudes for our [3.6] and [4.5] photometry of DENIS 1058
and five stars in the same IRAC field (shown as open squares in Figure \ref{fig:FAP}).

Simulations we have performed, as well as actual experience with other
WoW targets, indicate that pixel phase correction using Equation
\ref{eq:ppc01} is very unlikely to eliminate genuine astrophysical
variations, although it can distort them.  As shown in Figure \ref{fig:ch1raw}, 
Spitzer's pointing shows both a long-term, approximately linear trend and an oscillation
at a frequency of 1-1.5 cycles/hour.  The extent to which
the pixel phase correction can distort true astrophysical variability
depends on how the timescale of the astrophysical variability compares
to that of the pointing variations.  The periodogram peak for DENIS 1058
is at a period of 5.02 hours, which is shorter than the observation
interval but substantially longer than the pointing oscillation trend,
so distortion should not be severe.  However, as outlined above, we will
obtain our final result from an MCMC analysis that is less prone to distortion and
that allows better characterization of uncertainties than the periodogram analysis
we have used simply to demonstrate that DENIS 1058 is a true variable.
In fact, our MCMC analysis in \S~\ref{sec:ch1m} yields a period of
$4.25^{+0.26}_{-0.16}$ hours, which is not consistent with the 5.02-hour periodogram peak.
As illustrated
in Figure \ref{fig:pgram}, we have demonstrated using synthetic data that the
discrepancy is indeed due to the slight distortion of the astrophysical
variations that is imposed by the pixel phase correction.  The $\sim 4.25$ hour
value, based on the more sophisticated fits, is the correct one.

\begin{deluxetable}{ccccc}
\tablewidth{0pc}
\tablecolumns{5}
\tablecaption{Periodogram FAP Values for IRAC Data on DENIS 1058 and Field Stars \label{tab:pgram01}}
\tablehead{\colhead{Object} & \colhead{Magnitude [3.6]}  & \colhead{Magnitude [4.5]}  & \colhead{FAP [3.6]} & \colhead{FAP [4.5]} }
\startdata
DENIS 1058 & 11.76 & 11.76 & $2.63 \times 10^{-10}$ & $6.02 \times 10^{-1}$ \\
Star A & 12.81 & 12.84 & $4.62 \times 10^{-2}$ & $1.94 \times 10^{-1}$ \\
Star B & 13.23 & 13.18 & $6.07 \times 10^{-2}$ & $9.67 \times 10^{-1}$ \\
Star C & 13.11 & 13.12 & $2.52 \times 10^{-3}$ & $6.74 \times 10^{-1}$ \\
Star D & 13.21 & 13.14 & $5.02 \times 10^{-1}$ & $3.36 \times 10^{-1}$ \\
Star E & 13.42 & 13.50 & $4.41 \times 10^{-1}$ & $7.54 \times 10^{-1}$ \\
\enddata
\end{deluxetable}

\subsection{SVD Fits to Determine Input Parameters for MCMC Analysis} \label{sec:ch1s}

Our final analysis of DENIS 1058's variability will use an MCMC, 
but to determine the correct input model for
such an analysis and to estimate the uncertainty on our data points, 
we first perform least-squares fits to the data using
SVD.  Our astrophysical model is a truncated Fourier series:

\begin{equation}
h(t) = 1 + \sum_{i=1}^{n} C_i \cos \left( \frac{2 \pi t}{P/i} \right) +  S_i \sin \left( \frac{2 \pi t}{P/i} \right)
\label{eq:svd04}
\end{equation}

Given a fixed period $P$ and photometry already corrected for the pixel
phase effect, Equation \ref{eq:svd04} is linear in the parameters (the $C_i$ and $S_i$), 
so a least-squares solution could be obtained using SVD; a range of
periods could be tried and the one producing the lowest residual RMS identified.

However, in order to avoid distorting the astrophysical variability, we must
solve simultaneously for the pixel phase parameters
of Equation \ref{eq:ppc01}, which requires fitting the equation:

\begin{equation}
g(x,y,t) = f(x,y) h(t),
\label{eq:svd02}
\end{equation}

where $f(x,y)$ is the pixel phase function given in Equation \ref{eq:ppc01}.
Equation \ref{eq:svd02} is nonlinear due to the
multiplication of $f(x,y)$ and $h(t)$, and cannot be linearized
by taking a logarithm, because both the multiplied terms are themselves
the sums of independent functions.

However, since we normalize our data prior to the fit,
and since the amplitudes of both systematic and astrophysical terms
are small, Equation \ref{eq:svd02} can be approximated by $g(x,y,t) = f(x,y) + h(t) - 1$, 
which is linear in the parameters and thus can be solved using SVD.
We take the resulting approximations for  $f(x,y)$ and $h(t)$ as the starting 
point for an iterative solution of Equation \ref{eq:svd02}.
The first iteration begins with dividing the normalized raw photometry
by the approximate value for $f(x,y)$, which produces photometry approximately
corrected for the pixel phase effect.  We fit this photometry using Equation \ref{eq:svd04} to yield
an improved approximation of $h(t)$.  We divide the raw fluxes by this, and apply 
Equation \ref{eq:ppc01} to the resulting
photometry to obtain an improved solution for $f(x,y)$ --- which forms the starting
point of a new iteration.  This process converges rapidly 
even on eclipsing variables with astrophysical amplitudes greater than 20\%.  
Note that a separate iterative solution is obtained for each
period in a finely-sampled range, and the final output corresponds to the period that
yielded the lowest residual RMS.  Parameter values obtained by solving the linear, approximate version of
Equation \ref{eq:svd02} are always quite close to the final results: however, the residual
RMS is lower for the iterative solution of the full, physically self-consistent equation.

\begin{figure}
\includegraphics[scale=0.8]{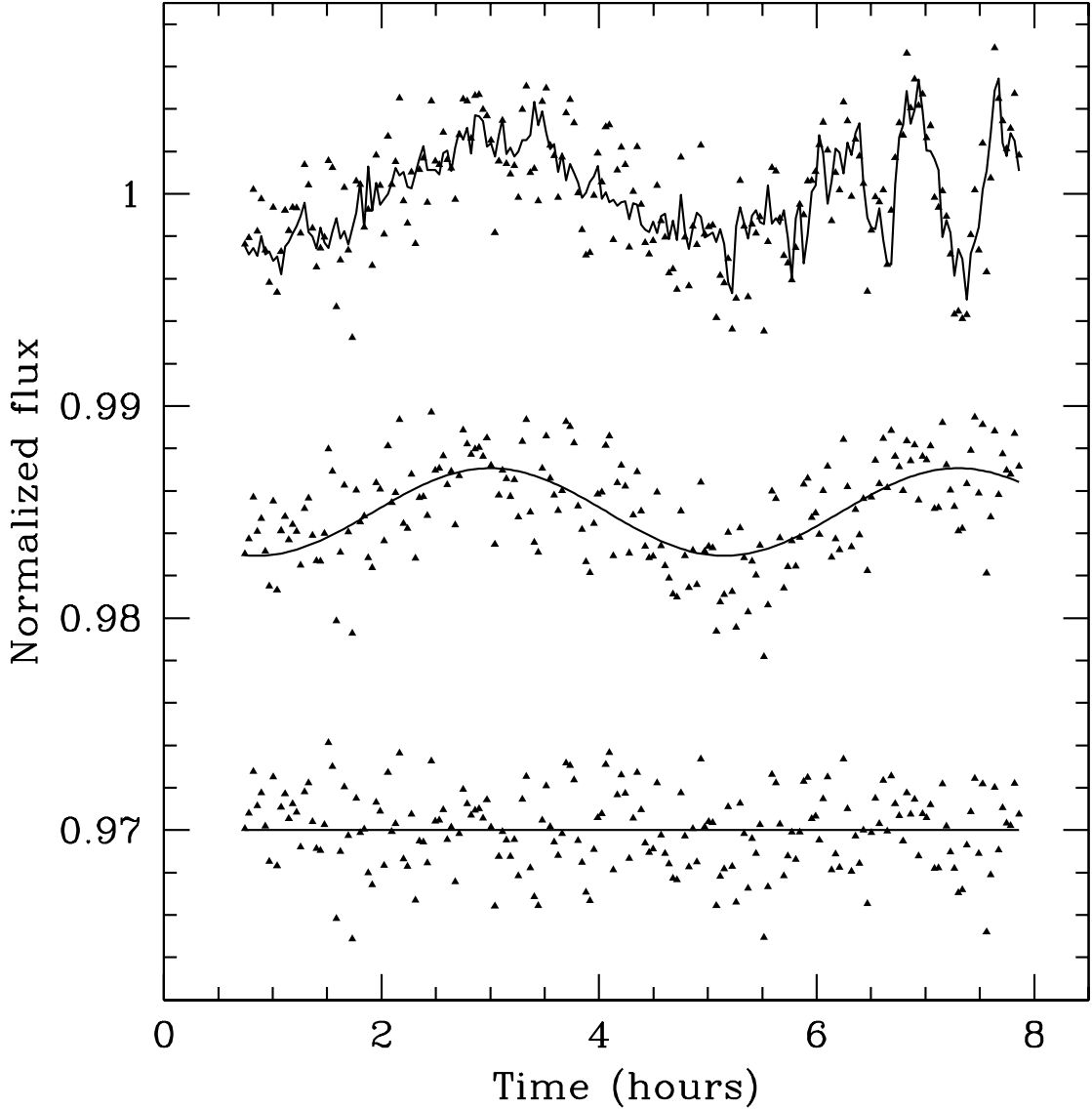}
\caption{\textbf{Top:} IRAC [3.6] photometry with a simultaneous fit
(Equation \ref{eq:svd02}) to both the pixel phase effect
and a sinusoidal model of the astrophysical variability.
\textbf{Middle:} Photometry after correction for the pixel
phase component of the fit (division by $f(x,y)$), with
the best-fit astrophysical model $h(t)$. This sinusoidal model has a period
of 4.23 hours and a peak-to-valley amplitude of 0.393\%.
\textbf{Bottom:} Residuals from the full fit, which are
consistent with random noise.  For clarity, the corrected
photometry and the residuals are offset vertically by -0.015 and
-0.03, respectively.
\label{fig:ffit01}}
\end{figure}

We determine the best number $n$ of Fourier terms for fitting a given
data set by performing a Lomb-Scargle periodogram analysis of the residuals
from the fit.  We choose the lowest value of $n$ that yields a FAP for
the residuals that is greater than $10^{-2}$, indicating that all
measurable astrophysical variations have been successfully modeled.
For DENIS 1058, fitting with only one Fourier term yields a residual
FAP of 0.35, demonstrating that a pure sinusoid is a sufficient model.  
The sinusoid we obtain by solving Equation 
\ref{eq:svd02} has a period of 4.23 hours and a peak-to-peak amplitude
of 0.393\%.  The residual RMS from this fit is 0.186\%, which is identical
to the value obtained by dividing the measured single-point RMS (Table \ref{tab:data})
by $\sqrt 10$ to account for the 10-point binning used in our analysis.
This agreement indicates that all astrophysical and systematic terms have been
effectively modeled.
Figure \ref{fig:ffit01} shows the full
solution to Equation \ref{eq:svd02}, together with the sinusoidal model $h(t)$ and
the final residuals.

We confirm the robustness of this solution by re-solving with different
initial trims, and with a cubic rather than quadratic version of the
pixel-phase function $f(x,y)$.  Initial trims from 10-40 minutes under quadratic
correction, and 0.6-40 minutes under cubic correction, produce periods
and amplitudes spanning relatively narrow ranges of 4.21 to 4.44 hours
and 0.388 to 0.413\%, with the residual RMS somewhat elevated
under the least aggressive trims.  


We have also experimented with photometric apertures that vary according to
the value of the noise pixel parameter $\tilde \beta$, which measures
the width of the instrumental PSF (see Lewis et al. 2012).
We find that such variable apertures produce markedly poorer photometry.  
Simliarly, including linear and quadratic terms dependent on $\tilde \beta$
in our fit (i.e. turning $f(x,y)$ into $f(x,y,\tilde \beta)$) produces
only an insignificant reduction in
the standard deviation of residuals (0.186\% to 0.184\%), without changing
the astrophysical parameters to any substantive degree.  It is not surprising
that the optimal photometric analysis for DENIS 1058 would differ from that
for a much brighter object such as is analyzed by \citet{np}.

\subsection{MCMC Analysis} \label{sec:ch1m}

We subject our data to an MCMC analysis, using a nine-parameter model equivalent
to Equation \ref{eq:svd02}. We use a constant uncertainty of 0.186\% for
all the data points, equal to the RMS scatter from the best SVD fit. Following
\citet{MCMC01}, we allow only one, randomly selected parameter to change at
each link of the Markov chain, and we change it by a random amount distributed
according to a Gaussian of mean zero and standard deviation $\beta_{\mu}$.  Here 
$\mu$ indexes the nine fit parameters, and the $\beta_{\mu}$ values must be
set appropriately before the launch of the Markov chain.  We adjust them so that
20-50\% of the new trial values for each given parameter yield $\chi^2$ 
low enough to be accepted as a new link in the Markov chain.  We have confirmed
that the MCMC results are robust under different values of the $\beta_{\mu}$.
They are also robust under binning schemes different from our default 10-point
binning: MCMC analyses with unbinned data (12 second sampling) 
and 25-point binned data (5 minute sampling) produced results in agreement to
well within 1$\sigma$.

We use $2 \times 10^9$ realizations for our final MCMC analysis.  Following
\citet{MCMC01}, we discard the first 10\% of the Markov chain to prevent
biasing the final results by not-yet-converged early solutions.  We find
a period of $4.25^{+0.26}_{-0.16}$ hours and a peak-to-valley amplitude
of $0.388 \pm 0.043$\%, where 1$\sigma$ uncertainties are quoted based on the
unweighted distribution of the respective parameters over all solutions
accepted as links in the last 90\% of the Markov chain.  

\section{Spitzer IRAC [4.5] Data Analysis} \label{sec:ch2}

Up through the screening for astrophysical variability, the analysis
of our [4.5] data proceeds almost exactly as that for the [3.6] data
already described.  A slightly smaller optimal photometric aperture radius
(1.9 pixels) is found at [4.5], perhaps because the background noise
makes a larger contribution.  Because Spitzer made
only a very short slew from its [3.6] pointing position, there
is no pointing anomaly and no settling time at the beginning of
the [4.5] data: we trim only the very first frame, which is deviant in all IRAC
data sequences.  

\begin{figure}
\includegraphics[scale=0.8]{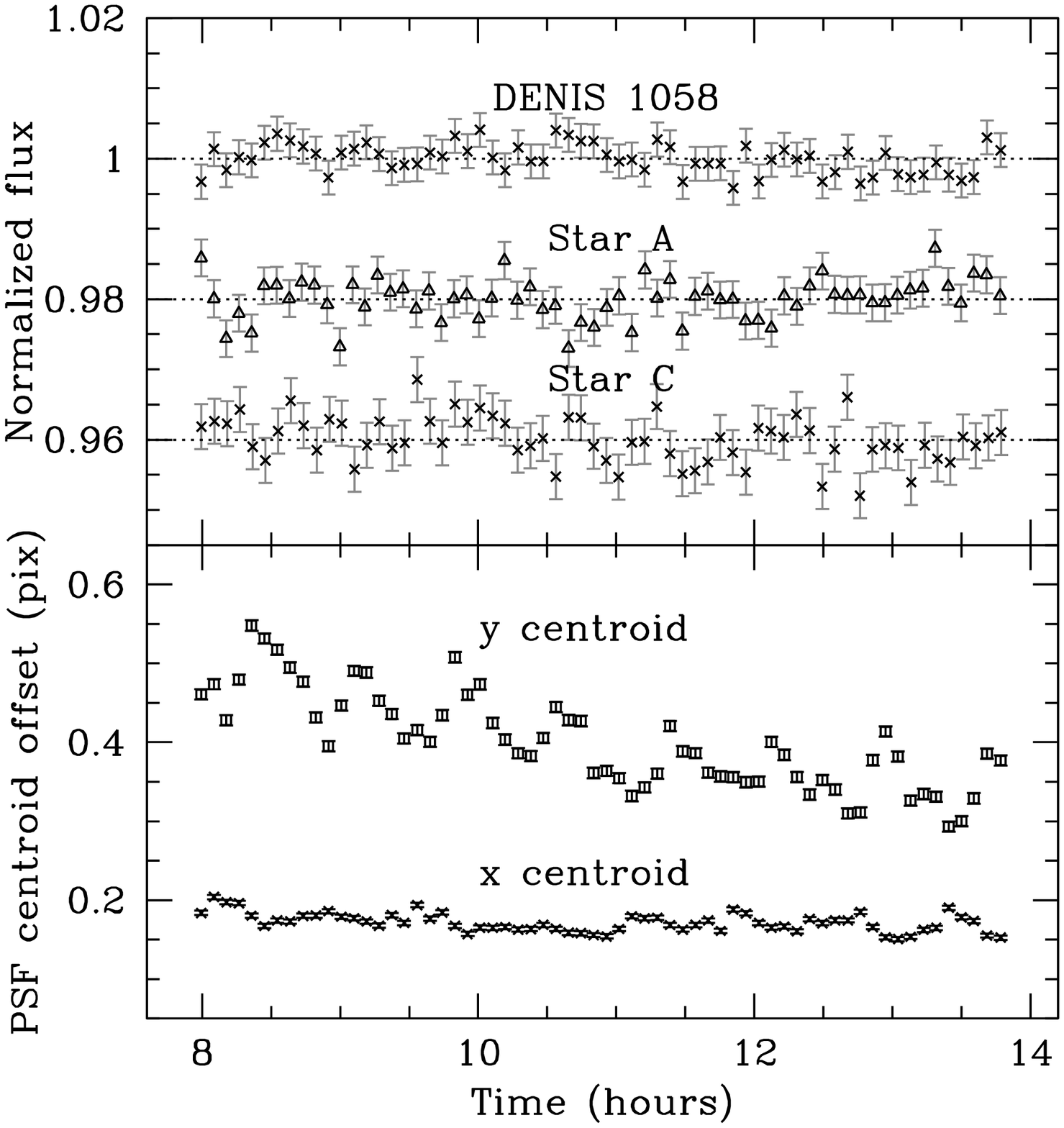}
\caption{\textit{Upper Panel}:
Normalized uncorrected IRAC [4.5] photometry of DENIS 1058 and two comparison
stars.  For clarity, the data are binned in sets of 25 points (each bin
thus represents a 5-minute time interval) and the photometry for Star A and Star C
is offset by -0.02 and -0.04, respectively, relative to DENIS 1058.
\textit{Lower Panel}: Location of the x and y centroids of DENIS 1058 
in the images.  An offset of 0,0 corresponds to the center of pixel 23,231 on the
IRAC [4.5] detector.  Note that there is no initial pointing anomaly analogous
to that seen in the [3.6] data. \label{fig:ch2raw}}
\end{figure}

\begin{figure}
\includegraphics[scale=0.8]{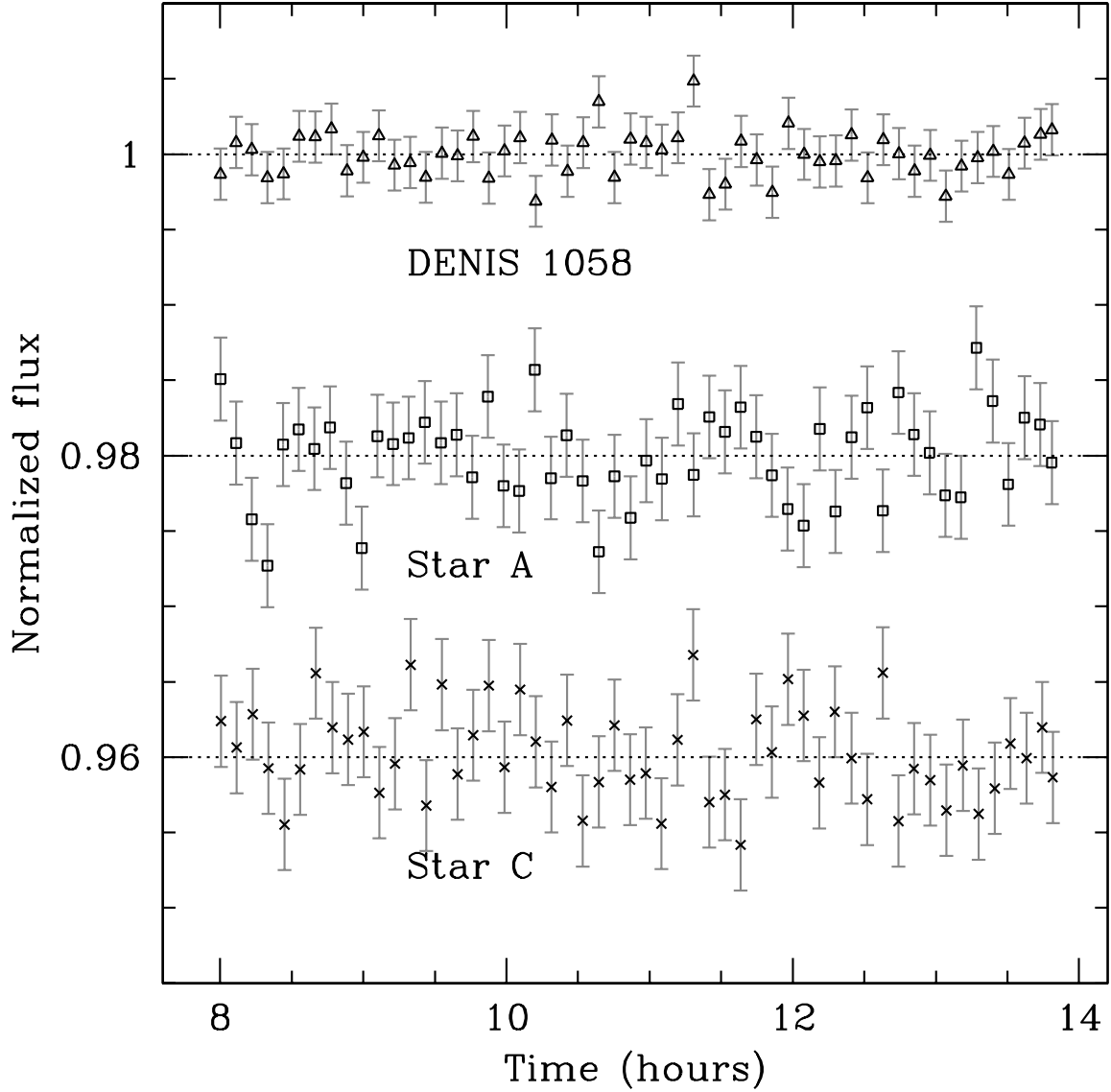}
\caption{Binned, normalized
IRAC [4.5] photometry of DENIS 1058 and the two brightest stars in the same IRAC field, after
correction for the `pixel phase' effect based on Equation \ref{eq:ppc01}.  For clarity, the points here
are 30-point (6-minute) bins relative to the raw data, and
photometry for Stars A and C is offset by -0.015 and -0.03, respectively.  Weak pixel-phase
artifacts visible in Figure \ref{fig:ch2raw} are well-corrected, but in
contrast to the [3.6] results, DENIS 1058 shows no evidence for variability
in [4.5]. \label{fig:ch2ppc}}
\end{figure}

Figure \ref{fig:ch2raw} shows our raw binned data.  Note that the time
series is contiguous with that of the [3.6] data, as the WoW
observations of each target are sequential.  
The pixel phase effect is weaker in [4.5] vs. [3.6], but correction
is still warranted.  Figure \ref{fig:ch2ppc} shows the data after
correction using Equation \ref{eq:ppc01}.

The FAP values of our [4.5] data are given in Table \ref{tab:pgram01}.
With two of five measured field stars showing lower FAP, there is no evidence
that DENIS 1058 exhibits significant variability at [4.5].  We note that
this conclusion is unaffected by the fact that thanks to weaker systematics and 
a shorter data sequence, FAP values tend to be higher for all objects at [4.5]
vs. [3.6]. 
As we will demonstrate below, this non-detection is not due to lower sensitivity
at the longer passband: variability with the same amplitude as at [3.6] would
easily be detected in our [4.5] data.

We emphasize that DENIS 1058 is variable at [3.6] beyond reasonable doubt,
as previously demonstrated by Figure \ref{fig:FAP}.  We note that
\citet{SpitzerPhot1} also found two L dwarfs that showed possible variations in one Spitzer IRAC
band but not in another (the bands were [4.5] and [8.0], respectively).  Because of the
lack of confirmation at [8.0], they refrained from claiming their [4.5] detection as
true astrophysical variability.  While such caution was warranted then, our own [3.6] detection
is confirmed by a systematic analysis of field stars that was beyond the scope of
previous work.  Since our result shows that the variability
amplitude of an L dwarf can be very different from one band to another, it may
suggest that the variations reported by \citet{SpitzerPhot1} at [4.5] had
a genuine astrophysical origin.

Our periodogram analysis suggests that our [4.5] photometry of DENIS 1058 is
consistent with zero variability, and Figures \ref{fig:ch2raw} and \ref{fig:ch2ppc}
support this.  An MCMC analysis of the same form as we used for the [3.6] data would thus
find sinusoidal amplitudes consistent with zero and therefore fail to converge on
meaningful values for the period and phase.  To avoid this, we perform an MCMC
analysis with the period and phase fixed to the final values from [3.6] analysis.
Thus our [4.5] MCMC analysis has seven rather than nine parameters: the six $A_i$ from
Equation \ref{eq:ppc01} plus only the amplitude of the sinusoid.

\begin{figure}
\includegraphics[scale=0.8]{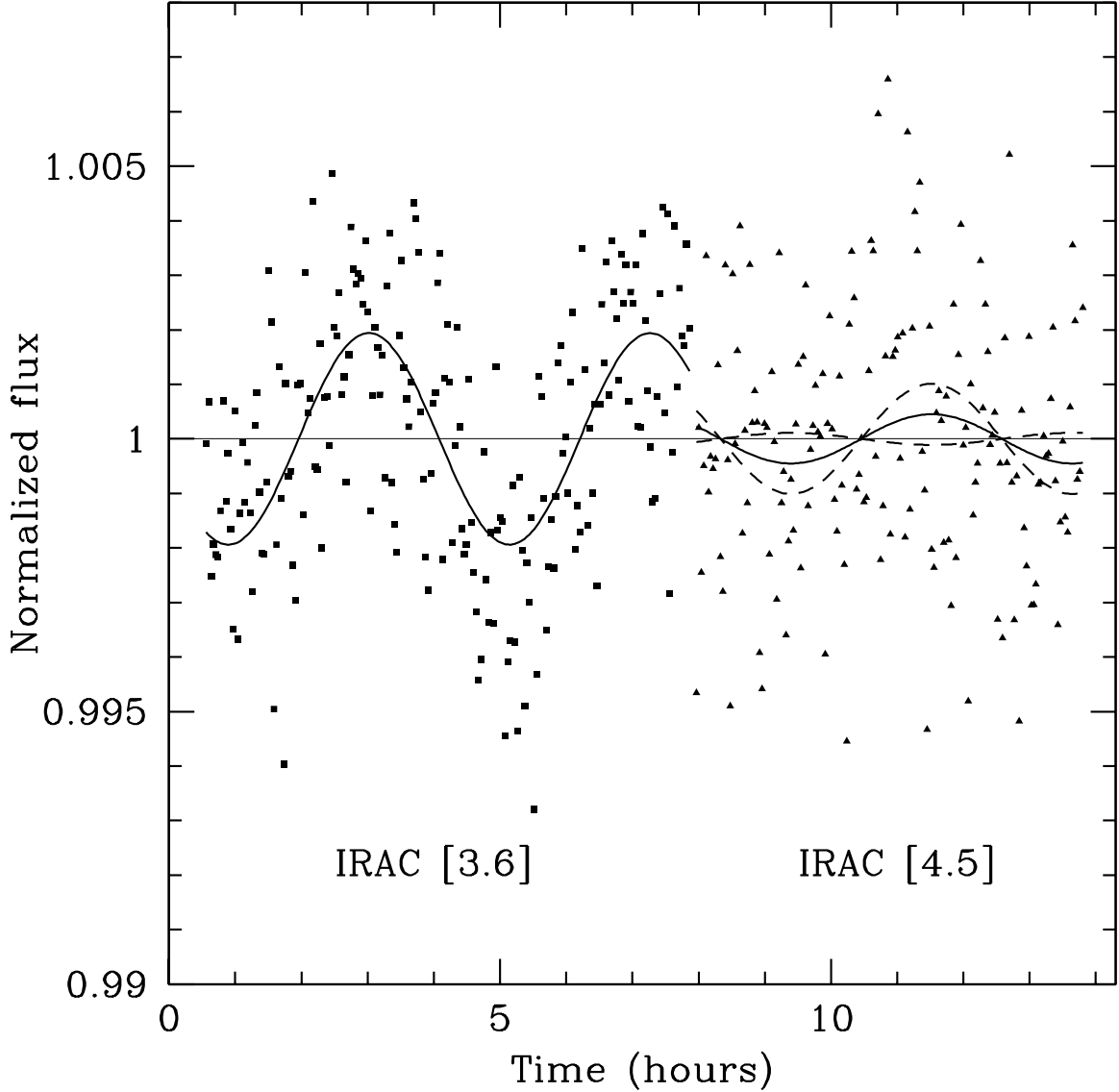}
\caption{IRAC [3.6] and [4.5] data after removal of the
pixel phase systematics based on simultaneous
fits to the pixel phase effect and a sinusoid.
The best fit sinusoids for both bands are shown as
solid lines, where the [4.5] curve has been constrained
to have the same period and phase as that at [3.6].
The dashed curves give the minimum and maximum amplitudes
permitted for the [4.5] data at the 2$\sigma$ level.
The former amplitude is negative and corresponds
to extremely low-amplitude variations anti-correlated with those observed at [3.6],
but demonstrates that the data are also consistent with zero
variability at [4.5].  The data suggest weak positively
correlated variability, but zero or anti-correlated variability
cannot be ruled out.
\label{fig:2chan}}
\end{figure}

This analysis yields a peak-to-valley amplitude of $0.090 \pm 0.056$\%,
which corresponds to a [4.5]/[3.6] amplitude ratio of $0.23 \pm 0.15$.
While the positive amplitude ratio suggests that DENIS 1058 exhibits weak [4.5]
variability positively correlated with that at [3.6], we note that the data
are consistent with zero and even negative amplitudes (corresponding to
anti-correlated variations).  By contrast, [4.5] variations
with amplitude equal to those at [3.6] are excluded at the 5$\sigma$ level.
Figure \ref{fig:2chan} combines the data and best-fit
sinusoids for both [3.6] and [4.5], illustrating the striking difference
in DENIS 1058's photometric behavior in the two bands.

\section{SOAR $J$-band Analysis} \label{sec:$J$-band}

We begin our $J$-band analysis of DENIS 1058 by identifying 14 field stars bright
enough to be potentially useful for relative photometry.
They are shown in Figure \ref{fig:Jfield} and listed in Table \ref{tab:Jref}.

\begin{deluxetable}{cccccccc}
\tablewidth{0pc}
\tablecolumns{8}
\tablecaption{Objects measured in SOAR $J$-band images \label{tab:Jref}}
\tablehead{\colhead{{\footnotesize Designation}} & \colhead{{\footnotesize Catalog}} & \multicolumn{2}{c}{{\footnotesize J2000.0 Coordinates}} & \colhead{{\footnotesize 2MASS $J$}} & \colhead{{\footnotesize 2MASS $J-K_S$}} & & \colhead{{\footnotesize SOAR}} \\
\colhead{{\footnotesize (this work)}} & \colhead{{\footnotesize Designation\tablenotemark{a}}} &\colhead{{\footnotesize RA}} & \colhead{{\footnotesize DEC}} & (mag) & (mag) & \colhead{{\footnotesize $F$\tablenotemark{b}}} & \colhead{{\footnotesize relative RMS\tablenotemark{c}}}} 
\startdata
{\footnotesize DENIS 1058} & {\footnotesize 2MASS J10584787-1548172} & {\footnotesize 10:58:47.87} & {\footnotesize -15:48:17.2} & {\footnotesize $14.16\pm0.04$} & {\footnotesize $1.62 \pm 0.05$} & {\footnotesize 20.11} & {\footnotesize 1.07\%} \\
{\footnotesize Star 2} & {\footnotesize 2MASS J10584625-1548513} & {\footnotesize 10:58:46.26} & {\footnotesize -15:48:51.4} & {\footnotesize $13.59\pm0.03$} & {\footnotesize $0.43 \pm 0.05$} & {\footnotesize 14.43}  & {\footnotesize 0.72\%} \\
{\footnotesize Star 3} & {\footnotesize 2MASS J10584730-1548500} & {\footnotesize 10:58:47.30} & {\footnotesize -15:48:50.0} & {\footnotesize $13.88 \pm 0.03$} & {\footnotesize $0.76 \pm 0.05$} & {\footnotesize 15.23}  & {\footnotesize 0.54\%} \\
{\footnotesize Star 4} & {\footnotesize 2MASS J10584309-1548310} & {\footnotesize 10:58:43.09} & {\footnotesize -15:48:31.1} & {\footnotesize $14.81 \pm 0.04$} & {\footnotesize $0.45 \pm 0.11$} & {\footnotesize 15.63}  & {\footnotesize 1.18\%} \\
{\footnotesize Star 5} & {\footnotesize 2MASS J10583944-1550392} & {\footnotesize 10:58:39.44} & {\footnotesize -15:50:39.3} & {\footnotesize $16.06 \pm 0.09$} & {\footnotesize $0.60 \pm 0.25$} & {\footnotesize 17.30}  & {\footnotesize 1.37\%} \\
{\footnotesize Star 6} & {\footnotesize 2MASS J10583674-1550562} & {\footnotesize 10:58:36.74} & {\footnotesize -15:50:56.3} & {\footnotesize $15.95 \pm 0.07$} & {\footnotesize $0.85 \pm 0.18$} & {\footnotesize 17.01}  & {\footnotesize 0.89\%} \\
{\footnotesize Star 7} & {\footnotesize 2MASS J10583905-1549451} & {\footnotesize 10:58:39.05} & {\footnotesize -15:49:45.2} & {\footnotesize $12.87 \pm 0.02$} & {\footnotesize $0.80 \pm 0.04$} & {\footnotesize 14.46}  & {\footnotesize 0.85\%} \\
{\footnotesize Star 8} & {\footnotesize 2MASS J10582955-1547299} & {\footnotesize 10:58:29.55} & {\footnotesize -15:47:29.9} & {\footnotesize $16.23 \pm 0.10$} & {\footnotesize $\sim 0.77$\tablenotemark{d}} & {\footnotesize 16.77}  & {\footnotesize 1.31\%} \\
{\footnotesize Star 9} & {\footnotesize 2MASS J10583882-1546204} & {\footnotesize 10:58:38.83} & {\footnotesize -15:46:20.5} & {\footnotesize $12.56 \pm 0.02$} & {\footnotesize $0.41 \pm 0.03$} & {\footnotesize 13.46}  & {\footnotesize 1.03\%} \\
{\footnotesize Star 10} & {\footnotesize 2MASS J10584149-1547206} & {\footnotesize 10:58:41.50} & {\footnotesize -15:47:20.7} & {\footnotesize $16.16 \pm 0.10$} & {\footnotesize $0.64 \pm 0.25$} & {\footnotesize 17.23}  & {\footnotesize 1.31\%} \\
{\footnotesize Star 11} & {\footnotesize GSC 2.3 S5IT005181} & {\footnotesize 10:58:40.78} & {\footnotesize -15:45:53.7} & {\footnotesize \nodata} & {\footnotesize \nodata} & {\footnotesize 18.56}  & {\footnotesize 2.14\%} \\
{\footnotesize Star 12} & {\footnotesize GSC 2.3 S5IT005141} & {\footnotesize 10:58:37.94} & {\footnotesize -15:46:08.8} & {\footnotesize \nodata} & {\footnotesize \nodata} & {\footnotesize 18.50}  & {\footnotesize 2.80\%}  \\
{\footnotesize Star 13} & {\footnotesize GSC 2.3 S5IT005115} & {\footnotesize 10:58:41.32} & {\footnotesize -15:46:15.3} & {\footnotesize \nodata} & {\footnotesize \nodata} & {\footnotesize 19.64}  & {\footnotesize 4.14\%}  \\
{\footnotesize Star 14} & {\footnotesize GSC 2.3 S5IT005041} & {\footnotesize 10:58:38.22} & {\footnotesize -15:46:46.5} & {\footnotesize \nodata} & {\footnotesize \nodata} & {\footnotesize 18.23}  & {\footnotesize 3.87\%} \\
{\footnotesize Star 15} & {\footnotesize GSC 2.3 S5IT004646} & {\footnotesize 10:58:39.33} & {\footnotesize -15:48:59.6} & {\footnotesize \nodata} & {\footnotesize \nodata} & {\footnotesize 20.09}  & {\footnotesize 3.53\%} \\
\enddata
\tablenotetext{a}{\footnotesize{DENIS 1058 and Stars 2-10 were
found in the 2MASS catalog; the rest of the stars had no 2MASS detections
but were found in the GSC 2.3 catalog.}}
\tablenotetext{b}{\footnotesize{Red photographic magnitudes from the POSS.  Uncertainties
are typically 0.4-0.5 mag.}}
\tablenotetext{c}{\footnotesize{This is the RMS scatter of normalized
relative photometry of each star.  It was calculated before removal
of the suspected variable, Star 7.  Its chief value is simply as a rough
relative metric for the random scatter of each object.}}
\tablenotetext{d}{\footnotesize{There is a photometry error flag on the 2MASS K-band
magnitude of this object.  Its color is thus uncertain}}
\end{deluxetable}

We proceed to construct relative photometry of DENIS 1058 by ratioing its
flux on each image to the sum over measured fluxes of all fourteen reference
stars on the same image.  To screen for variability and explore the systematic
effects present in our data, we also construct relative photometry of the
reference stars by ratioing the flux of each to the summed flux of all the others:

\begin{equation}
R_{ji} =  \frac{F_{ji}}{\sum_{k=2, k \ne j}^{m} F_{ki}}
\label{eq:rp01}
\end{equation}

Here, $i$ indexes images while $j$ and $k$ index objects measured on each image, with $j = 1$ for
DENIS 1058 itself.  The $F_{ji}$ are instrumental fluxes while the $R_{ji}$ are the relative photometry.  
We optimize our photometric apertures to minimize the RMS scatter of the $R_{ji}$ 
for field stars with similar brightness
to DENIS 1058.  This results in an aperture of radius 13 pixels
(0.88 arcsec) for our in-focus images and 17 pixels (1.16 arcsec) for
the defocused images mentioned in \S~\ref{sec:Jdata}. For sky subtraction
we use an annulus of inner radius 55 pixels and width 10 pixels around each star.

\begin{figure}
\includegraphics[scale=0.8]{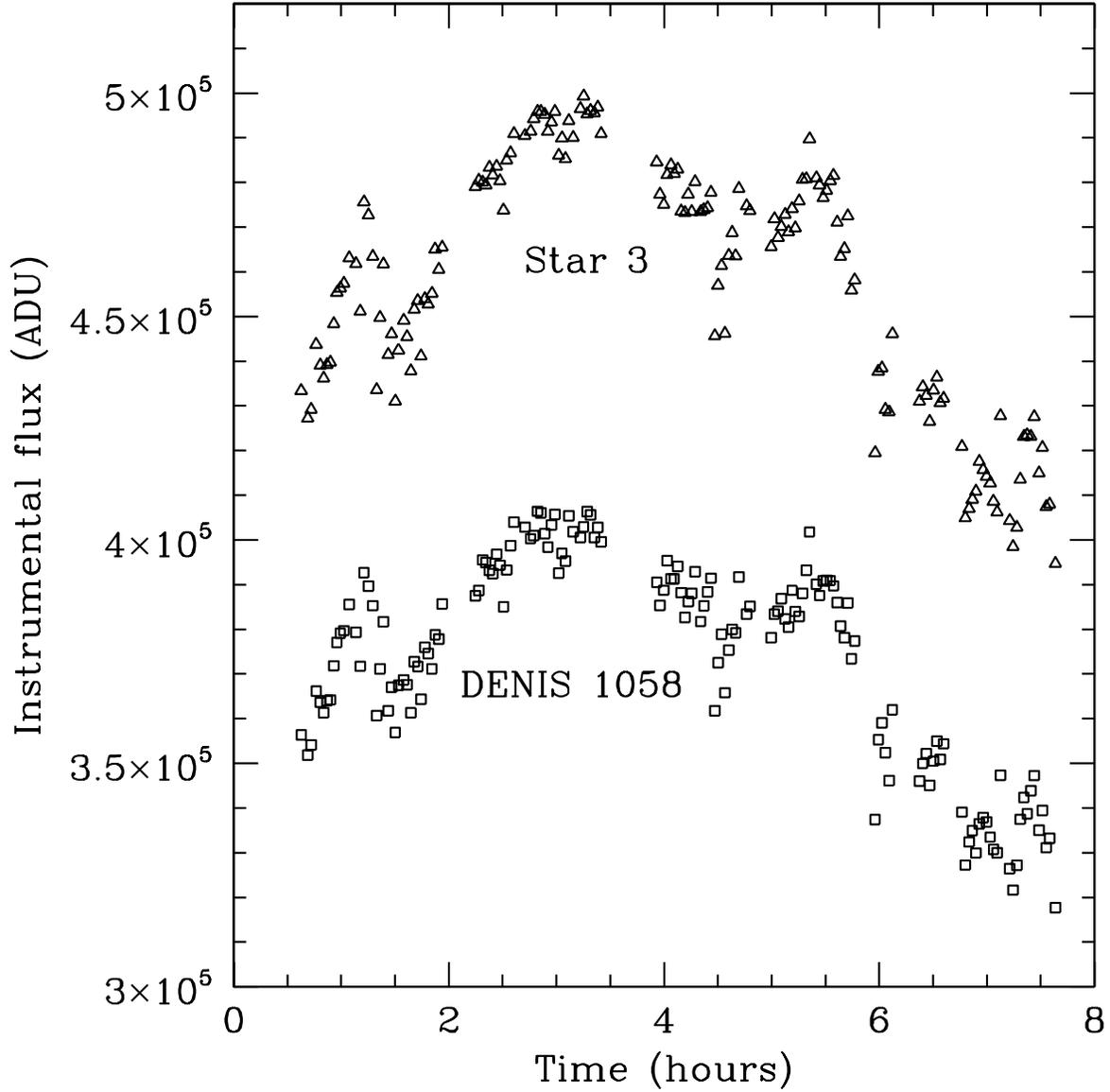}
\caption{Instrumental $J$-band photometry of DENIS 1058 and a similar-brightness field star from our
SOAR images. The $\sim$20\% range of variation seen here is mainly due
to variable aperture losses, which cancel out when relative
photometry is constructed by ratioing the flux of DENIS 1058 to the
summed fluxes of a set of non-variable field stars, including the
star shown here.\label{fig:rawJ}}
\end{figure}

Figure \ref{fig:rawJ} shows our raw photometry of DENIS 1058 and Star 3
($F_{1i}$ and $F_{3i}$) as a function of time.
The variations are caused predominately by changing aperture losses
due to seeing, telescope flexure, and focus adjustments.  Airmass
plays a secondary role, and both effects cancel out when the data are ratioed.

Our initial relative photometry shows deviant behavior for Star 7 and Star 9, the two
brightest objects in the field.  Further investigation shows that Star 9 occasionally
saturates, while Star 7 appears to be a variable. We reject both stars as photometric references.

\begin{figure}
\includegraphics[scale=0.8]{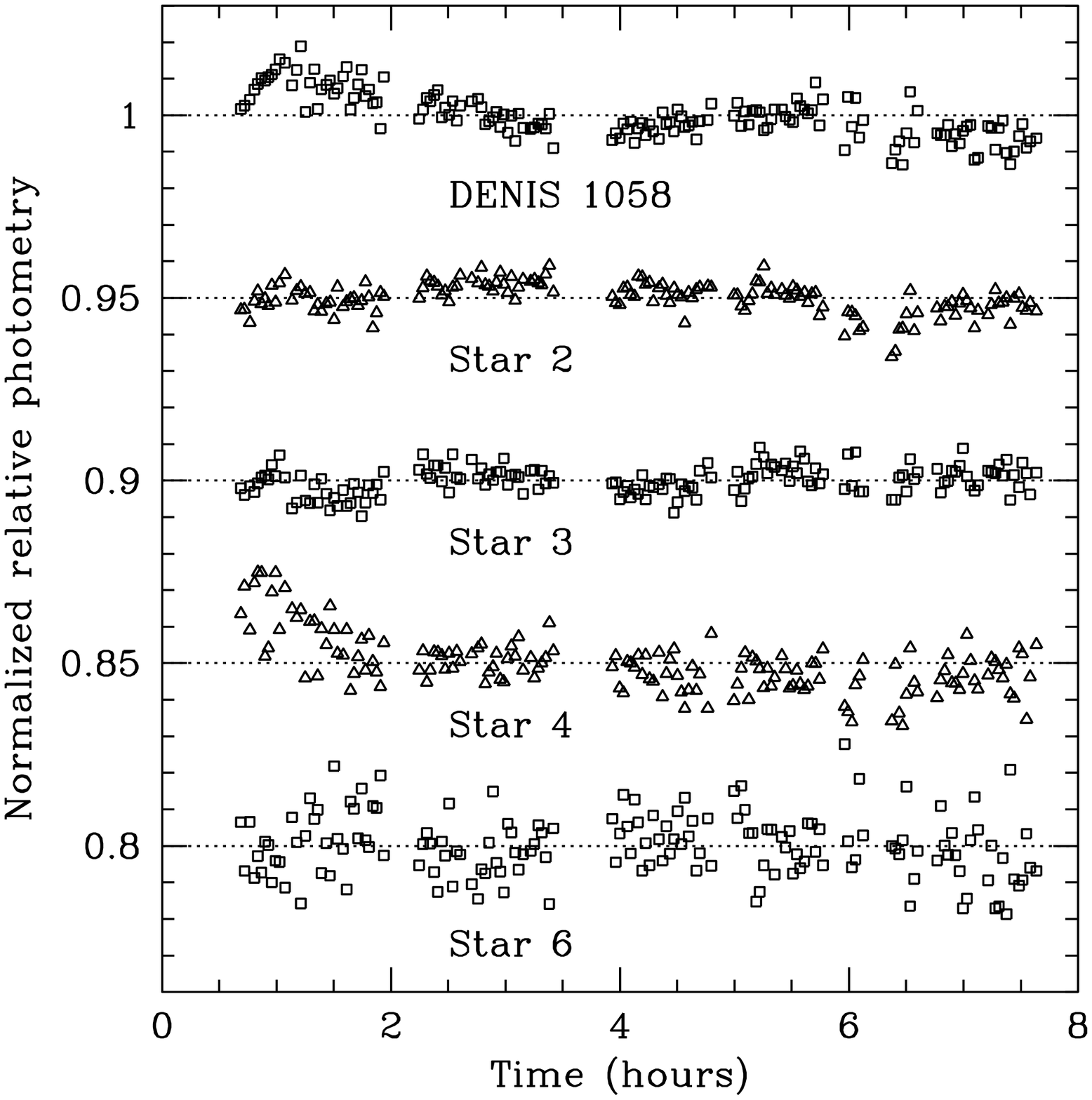}
\caption{Normalized relative photometry of DENIS 1058 and several
bright field stars in the $J$-band, after correction for the
photometric offset between the two nod positions. In the plot, we
have offset the data for each star by -0.05 relative to the previous
one, for clarity.  DENIS 1058 exhibits two well-defined local
extrema (at $\sim$3.8 and $\sim$5.7 hours) that are not at the endpoints
of the time series.  This property is consistent through
many different ways of fitting the data, and is not shared
by the field stars.  Data in the range 5.9-6.6 hours were taken
with the telescope purposely de-focused; the larger photometric aperture
required for these data increased the noise.
\label{fig:corJ01}}
\end{figure}

With these stars rejected, the largest remaining systematic variations
take the form of a clear bimodality in the relative fluxes from
images taken in one nod position vs. the other.  This is not unexpected due to
the likely existence of differing flatfield residuals at the two nod positions.  While
the amount of the offset differs from one star to another, it appears constant
in time for each star and is therefore easy to correct.  Because 
we nodded the telescope every $\sim240$ seconds during the data acquisition, the 
correction has no risk of distorting any but the highest frequency
astrophysical variations (e.g., asteroseismic pulsations) which, if present,
would have photometric amplitudes too small to be relevant here.  Figure
\ref{fig:corJ01} shows relative photometry of DENIS 1058 and
the four brightest non-variable field stars after correction for this
nod-offset effect.

DENIS 1058 already appears more variable than the other stars, with the
suggestion of a 4-5 hour sinusoidal variation consistent with the
IRAC [3.6] results (though at a larger amplitude). However, as three days elapsed 
between the SOAR and Spitzer observations, and the [3.6] period is not
sufficiently accurate to preserve phase information over this time interval,
we analyze the $J$-band data independently of the [3.6] results.  

Figure \ref{fig:corJ01} indicates some residual systematics in the
photometry of the field stars, as well as an apparent linear fading
trend superimposed on the approximately sinusoidal variations of DENIS 1058.
We have explored the origin of these systematics by multilinear SVD fits
to the normalized relative photometry of each object.  Beyond the systematic
offset between nod positions discussed above,
the systematic errors in our photometry show no clear correlation with
airmass, pixel position, or image sharpness.  Linear fits to relative
photometry as a function
of either airmass or time yield some reduction in the residual standard
deviation.  However, they do not
correct all systematic errors, and the linear time trend consistently 
produces a greater improvement than the
airmass or pixel-position fits.  A period near 4 hours and
an amplitude near 1\% is robustly found for DENIS 1058 under correction
by either a linear fit to airmass or a linear time trend; this fit is
also robust whether relative photometry is constructed by ratioing DENIS
1058 to the sum of all non-variable reference stars or only to the flux
of any single star among the three brightest (Stars 2-4).  While the systematic
variations in the field stars can be fit by sinusoids, no such sinusoidal fit approaches the consistency
of the fit to DENIS 1058 under different photometric ratios and selections
of systematic parameters.  

We emphasize that there is no evidence that any of the relative photometry
has a physically reasonable dependence on airmass.  In particular, despite
DENIS 1058's very different $J-K_S$ color relative to all the reference stars, the sign
of the airmass term is not consistent in fits to relative photometry constructed
by ratioing the L dwarf to different individual reference stars.
This is consistent with the fact that the bandpass of the MKO $J$ filter
used in the Spartan IR Camera does not
include wavelengths affected by strong telluric water-vapor absorption.
The earth's atmosphere has, in fact, almost a uniform opacity across
this band, which implies that objects of very different colors will
nonetheless experience identical airmass effects, consistent
with what we observe.  

We choose to model the systematics of our $J$-band photometry using two
parameters: the nod offset correction plus a linear time trend.
This is by far the best two-parameter model, and while it does not correct
all the systematic effects seen in the photometry of the field stars, we feel
that a more agressive choice would be too likely to distort the fit to DENIS 1058's
true astrophysical variability.

Since our model cannot remove all the systematic effects,
we created an iterative process to reduce them as much as possible,
obtaining improved relative photometry to input to our fits for both
systematic and astrophysical variations.  The objective of this process was
to prevent individual deviant points and systematic effects
specific to a particular star from affecting the relative photometry of the others.
The process is described in detail in Appendix A. 
We emphasize that it has no ability to remove either systematic errors or astrophysical
variations specific to a given object from that object's final photometry.
Its sole purpose is to reduce the effect that deviant photometry
\textit{of the reference objects} has on the relative photometry of
any given star.  Improvements to the photometry were subtle, 
but the occurrence rates of photometric outliers and
the standard deviations of fit residuals were reduced.  As a final
step, we fit a cubic polynomial in time to each object, and removed all
2.5$\sigma$ outliers from this fit.  This fit was for purposes
of trimming only and was not a correction applied to the data.
The maximum number of points clipped was 4 out of 159.

\begin{figure}
\includegraphics[scale=0.8]{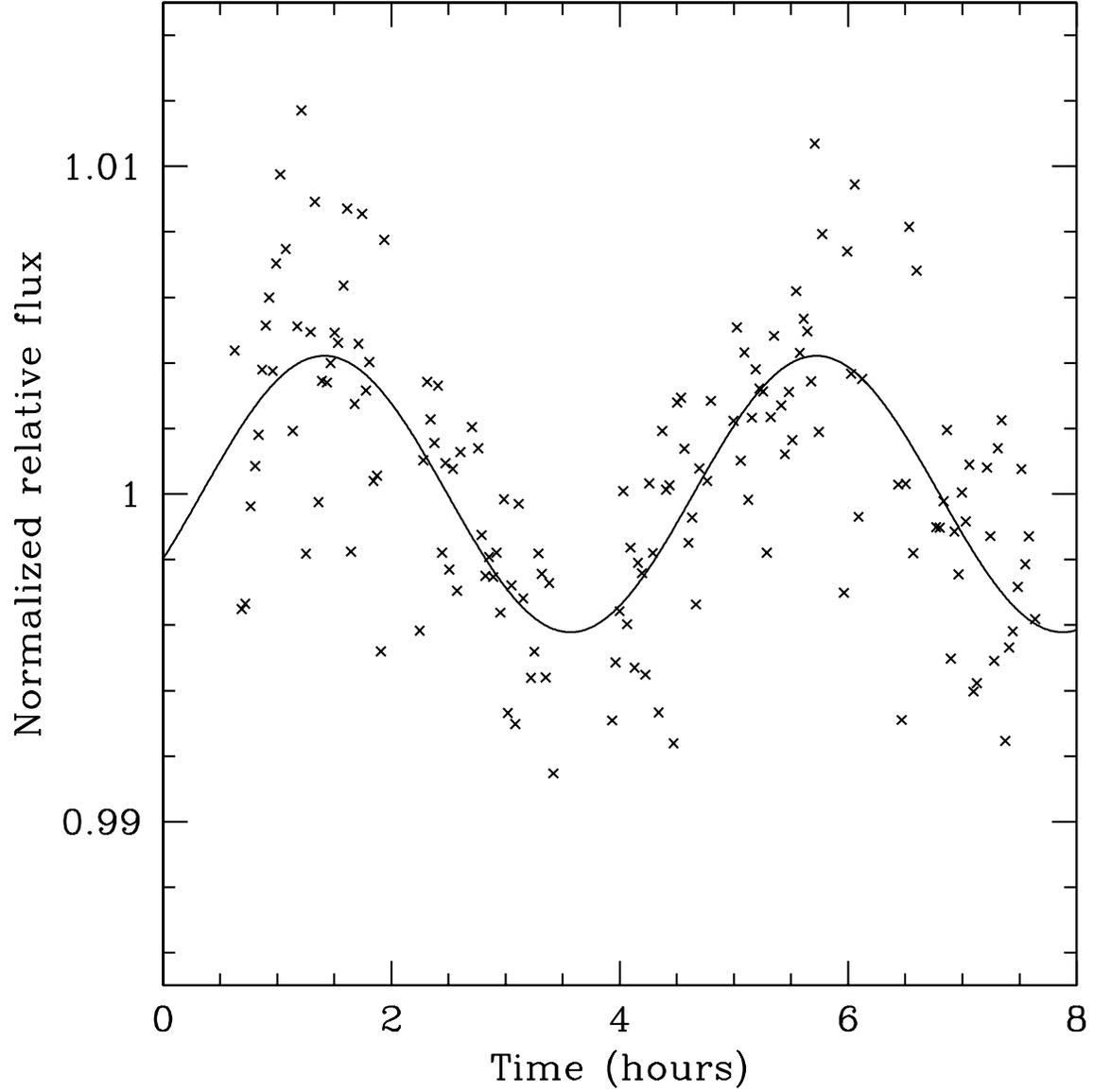}
\caption{Final normalized relative photometry of DENIS 1058 in the $J$-band, corrected
by the nod position offset and a linear time trend, and fit with the
best-fit sinusoid, having a period of 4.31 hours and a peak-to-valley
amplitude of 0.843\%.
\label{fig:Jonly}}
\end{figure}

We fit a sinusoid to our corrected $J$-band photometry of DENIS 1058
and each of the four brightest field stars,
using the same algorithm as for our IRAC [3.6] data to fit simultaneously
for the systematic error terms.  Consistent
with our initial results on the robustness of sinusoidal fits to DENIS 1058's
$J$-band photometry,
we find that although best-fit sinusoids do of course formally exist for all the
field stars, the sinusoidal component of the fit produces the most significant
reduction in the residual standard deviation for DENIS 1058.  The best-fit period
for DENIS 1058 in the $J$-band is $4.31 \pm 0.31$ hours, and the peak-to-valley
amplitude is $0.843 \pm 0.098$\%.  Figure \ref{fig:Jonly} shows this final fit.

Due to the presence of systematic effects for which we had no physical model,
quantifying the uncertainties we have quoted above required a different approach from
the MCMC analysis we applied to the IRAC data.  Instead, we used the
four brightest reference stars to obtain a sampling of the typical systematic errors,
imposed these errors on the photometry of DENIS 1058, and measured the resulting
scatter in the sinusoidal fit parameters. To do this, we modeled the normalized
photometry of each star using a cubic function of time and then multiplied
the relative photometry of DENIS 1058 by this function.  In this way we created four different
realizations of $J$-band photometry for DENIS 1058, each with the systematic errors of a different
reference star imposed.  We fitted the resulting distorted photometry using the
same method applied to the original data.  Thus we obtained five different
values for each parameter of the sinusoid: one from the original photometry
and one from each systematically altered version.  The uncertainties
quoted above are the standard errors from these five values.

Because of the systematics in our SOAR data and the
low amplitude of DENIS 1058's variability, if we did not have the IRAC [3.6] data as well,
we might report only a tentative detection of periodic $J$-band variability despite the
multiple lines of evidence in the $J$ band data that point to its reality.  Many such
tentative detections of L dwarf variability have been made, which are probably
real based on statistical arguments (e.g. Gelino et al. 2002; Koen 2003, 2004, and 2005).
However, given our IRAC [3.6] detection at a period matching the independently-derived $J$-band
period within 1$\sigma$, the reality of DENIS 1058's $J$-band periodic variability
is confirmed.


The amplitude of DENIS 1058's variability is higher in the $J$-band than in IRAC [3.6]
by a factor of $2.17 \pm 0.35$.  The fact that the amplitude is higher in the $J$-band relative
to longer wavelengths is consistent with theoretical models of cloud-induced
variability, and also matches observations of other variable brown dwarfs \citep{2M2139,Buenzli2012,Apai2013}.

\section{Astrophysical Implications} \label{sec:astrophysics}

DENIS 1058 varies in both the IRAC [3.6] and $J$-bands with
a consistent period but with substantially different amplitudes.
No significant variability is detected in IRAC [4.5], and any
variability in this band that is in phase with the [3.6] lightcurve
must (with 95\% confidence) have no more than about half the [3.6] amplitude. 
We now consider the origins of DENIS 1058's variability.

Periodic variability in stars is usually due to one of three types
of phenomena: close binaries and planetary companions (either eclipsing
or ellipsoidal variables), stellar pulsation, 
or rotation combined with magnetic star spots.  L dwarfs, being
cool enough to form condensate clouds, can also exhibit periodic variability 
due to rotation combined with inhomogenous cloud cover.

The significant differences in amplitude at different wavelengths in our
data suggest that the observed variability cannot be due to global changes
in DENIS 1058, such as those due to tidal effects from a close binary or
to pulsation.  \citet{pulsations} also find that the longest possible pulsational
periods for brown dwarfs are at least a factor of two shorter than the period
we have observed.  This leaves some type of rotational variability, induced either
by magnetic phenomena or inhomogenous clouds, as the preferred model.
In either case the variability has provided us with a rotation period.
Thus, before we consider the likely origin of the variability in more
detail, we first constrain the radius and age of DENIS 1058 based
on the \citet{Teff} measurement of its projected rotational velocity.

\subsection{Radius and Age} \label{sec:rad}

\citet{Teff} measure the projected rotational velocity of DENIS
1058 to be $v \sin(i) = 37.5 \pm 2.5$ km s$^{-1}$ based on line broadening in their 
Keck/HIRES spectrum.  Given rotation period $P$ and equatorial rotation velocity
$v$, an object's radius is $R = Pv/(2 \pi)$.  Since $v \sin(i)$ constitutes
a lower limit on the true rotational velocity, we can use it to get a lower limit
on the radius of DENIS 1058.  Using our IRAC [3.6] period of $4.25^{+0.26}_{-0.16}$ hours,
we find that $R = 0.131^{+0.012}_{-0.010} ~R_{\sun}$, which yields a 2$\sigma$ lower
limit of $0.111 ~R_{\sun}$.

This lower limit radius permits us to set upper limits on the age and mass of
DENIS 1058, using the fact that brown dwarfs contract over time and more massive ones
have smaller radii at a given temperature.  The only additional input we
need is $T_{\mathrm{eff}}$.  As reviewed in \S~\ref{sec:targ}, three analyses
have consistently found $T_{\mathrm{eff}} \sim 1950$ K. Only \citet{Dahn02}
quote an uncertainty, finding $T_{\mathrm{eff}} = 1945 \pm 65$ K.  Using the
evolutionary models of \citet{marley} for objects of solar metallicity
and $f_{sed} = 2$ (generally a good fit for L dwarfs; see Stephens et al. 2009),
we find that the largest-mass (and oldest) model consistent with our radius
limit has age 320 Myr, mass $0.055 ~M_{\sun}$, luminosity $1.90 \times 10^{-4} ~L_{\sun}$, 
$\log(g) = 5.09$, and $T_{\mathrm{eff}} = 2030$ K.

This model is, however, inconsistent with other data.  First, the luminosity
is much too high.  The results of \citet{spec01} and \citet{Dahn02} allow us
to calculate the bolometric luminosity of DENIS 1058 at $(1.03 \pm 0.07) \times 10^{-4} ~L_{\sun}$,
where we have set the uncertainty on the bolometric correction to 5\%.
This is inconsistent with the 320 Myr model by 12$\sigma$ (neglecting
uncertainties in the model luminosity).  
Secondly, a 320 Myr age is probably inconsistent with the lithium non-detections
reported by \citet{noLi}, \citet{martin97}, and \citet{Kirk}.
We suggest two possible
resolutions to the discrepancy, in the form of two parameters that, when all
available uncertainties are considered, show only a $\sim 3\sigma$ disagreement.

Firstly, we consider $T_{\mathrm{eff}}$.  Adopting our lower-limit radius, we can
use the measured luminosity and the equation
$T_{\mathrm{eff}} = \left(\frac{L}{4 \pi R^2 \sigma}\right)^{1/4}$ to find
$T_{\mathrm{eff}} = 1750 \pm 30$ K, which differs from the \citet{Dahn02}
result by only 2.7$\sigma$.  We note also that \citet{Dahn02} adopted a radius 
of $0.0903 ~R_{\sun}$ in their calculation, and that a cooler $T_{\mathrm{eff}}$
will apply if the true radius is larger.  \citet{martin99} and \citet{Teff}
report $T_{\mathrm{eff}} = 1900$ K and 1950 K, respectively, but the uncertainties on
these values may be large enough not to be inconsistent with $T_{\mathrm{eff}} = 1750 \pm 30$ K.
More problematically, however, an object with $T_{\mathrm{eff}} = 1750 \pm 30$ K
and $R = 0.111 ~R_{\sun}$ would have age and mass well below 320 Myr and $0.055 ~M_{\sun}$, exacerbating
the inconsistency with lithium non-detections.  

Secondly, therefore, we
consider the rotation speed.  \citet{noLi} comment that
a model with mass $0.065 ~M_{\sun}$ and age 800 Myr would be consistent with their
spectral data.  This statement still holds based on the \citet{marley} models, which indicate
that such an object would have $T_{\mathrm{eff}}$  and luminosity consistent
with measurements.  Its radius would
be $0.100 ~R_{\sun}$, which we can combine with our period to find a rotational
velocity of $28.6^{+1.1}_{-1.6}$ km s$^{-1}$.  This disagrees with the \citet{Teff}
value by only 3.3$\sigma$, and we note that this is without including in the error
propagation any estimate for the uncertainty on the theoretical radius.

Regardless of which (if either) of these possible resolutions
for the discrepancy is to be preferred, our large radius estimate
for DENIS 1058 demonstrates that our viewing geometry must be approximately
equator-on.  Changing the assumed inclination from $90^{\circ}$ to $45^{\circ}$,
for example, yields $R = 0.186 \pm 0.011 ~R_{\sun}$, which could be reconciled
with the measured luminosity only by adopting a $T_{\mathrm{eff}}$ 
of less than 1450 K.  Such a value would be inconsistent with the observed
spectral type and would also imply a very young, low-mass object that should
show prominent lithium absorption.  Thus, while the discrepancy described
above prevents us from placing a formally precise limit on the inclination of
DENIS 1058, a value of at least $45^{\circ}$ is strongly implied.

\subsection{Photospheric Spots and Clouds} \label{sec:mod}

In this section we consider inhomogenous clouds and/or magnetic starspots as possible
causes of the variability we observe in DENIS 1058.  Cool starspots are produced
when locally strong magnetic fields inhibit convective heat transport in the
stellar atmosphere.  As we discuss below in \S~\ref{sec:magspots}, they may
not be able to form in L dwarfs, but for purposes of the present analysis we
will grant them to be at least a possibility.  Warm spots could arise
from the deposition of magnetic energy in the photosphere (producing
continuum emission) or the chromosphere (producing line emission).  
There are no published observations of the former (that is, persistent photospheric
warm spots of probable magnetic origin); nevertheless we will consider the possibility
briefly in the current section. Variability due to magnetic line emission 
will be considered in \S~\ref{sec:magsemm}.

If DENIS 1058's variability is due to photospheric spots with a large
temperature differential, we would expect them to exhibit high surface brightness contrast
across a wide range in wavelengths.  This is inconsistent with the
large differences in observed variability amplitude between [4.5], [3.6], and the $J$-band.
For example, although the [3.6] and $J$-band variability of DENIS 1058 could
be explained by a photospheric warm spot with a $T = 2880 \pm 210$ K blackbody
spectrum\footnote{Magnitudes were converted to fluxes for use in this calculation 
(and that in \S \ref{sec:magsemm} below) based on information from the IRAC Instrument Handbook 
(http://irsa.ipac.caltech.edu/data/SPITZER/docs/irac/iracinstrumenthandbook) 
and from Cohen et al. 2003.}, this scenario
overpredicts the variability amplitude at [4.5] by more than 6$\sigma$.
More sophisticated modeling described below reaches the same conclusion:
neither cold nor hot spots, regardless of the temperature differential,
can explain the observed variability in the absence of inhomogenous clouds.

Following similar analyses performed on T dwarfs by \citet{SIMP0136}, \citet{2M2139},
and \citet{Apai2013}, we construct a two-phase model of DENIS 1058: a primary phase modeling
the expected global overcast, and a secondary phase with different temperature 
and/or cloud parameters.  For both phases we use different model spectra from \citet{marley}.
A spectral fit to establish the temperature and
cloud properties appropriate for the primary phase is
beyond the scope of this work.  Instead, we rely on existing analyses
that have consistently found $T_{\rm eff} \sim 1950\,\rm K$ for DENIS 1058, remembering
also that its $J-K_S$ color gives no indication of unusual atmospheric
properties.  Thus, for the primary phase we use models
with $T_{\rm eff} = 1950\,\rm K$, $\log(g) = 5.0$, and cloud parameters corresponding to moderately
thick clouds ($f_{\mathrm{sed}} =$ 1, 2, or 3), consistent with those that, e.g., \citet{stephens} have found 
to match L dwarfs with spectral types similar to DENIS 1058.  We note that most of the models
considered in \S ~\ref{sec:rad} had $\log(g)$ fairly close to 5.0.  If DENIS 1058
matches the young model with $T_{\rm eff} = 1750$ K, the spectral models we
consider in the currect section will be somewhat incorrect, but the basic
conclusions should still apply.

The \citet{marley} theoretical spectra that we consider for modeling the
secondary phase have $\log(g) = 5.0$, with $T_{\mathrm{eff}}$ values ranging
from 1500 to 2300 K in intervals of 100 K, 
and five different values for the cloudiness parameter $f_{\mathrm{sed}}$:
1, 2, 3, 4, and $\infty$.  The latter quantity parameterizes the extent to which sedimentation, or rain,
occurs for the clouds: thus the models with $f_{\mathrm{sed}} = 1$ have the thickest clouds (fewest cloud
particles removed by sedimentation) while $f_{\mathrm{sed}} = \infty$ corresponds to a completely cloudless
case.  We have interpolated logarithmically in $T_{\mathrm{eff}}$ to obtain 
models with a spacing of 10 K.  

We seek to match three observables: our IRAC [3.6] amplitude, our [4.5]/[3.6] amplitude
ratio, and our $J$/[3.6] amplitude ratio.
For a given primary phase/secondary phase pair, we integrate
the spectral models over the bandpasses of the IRAC [3.6], IRAC [4.5], and MKO $J$ filters
used in our observations to get fluxes in each filter for each model.  We then assume that one
side of DENIS 1058 is completely covered by the primary phase, while the other side has secondary-phase
regions extending over a fractional area $\epsilon$.  We solve for $\epsilon$ based
on the [3.6] amplitude.  Let $p$ be the primary-phase model flux integrated over the IRAC [3.6] band, 
and $s$ be the secondary-phase model integrated over the same band.
Then the amplitude of variation is:

\begin{equation}
A_{[3.6]} = \frac{(1-\epsilon) p + \epsilon s - p}{p}.
\label{eq:amod01}
\end{equation}

Here, the numerator is simply the flux from the hemisphere where the secondary phase
regions appear (which the equation implicitly assumes is the brighter side, provided the amplitude
is positive) minus the flux $p$ from the hemisphere uniformly covered by the primary phase.
Simplifying and solving for $\epsilon$, we find:

\begin{equation}
\epsilon = \frac{A_{[3.6]} p}{s-p}.
\label{eq:amod02}
\end{equation}

\begin{figure}
\includegraphics[scale=0.8]{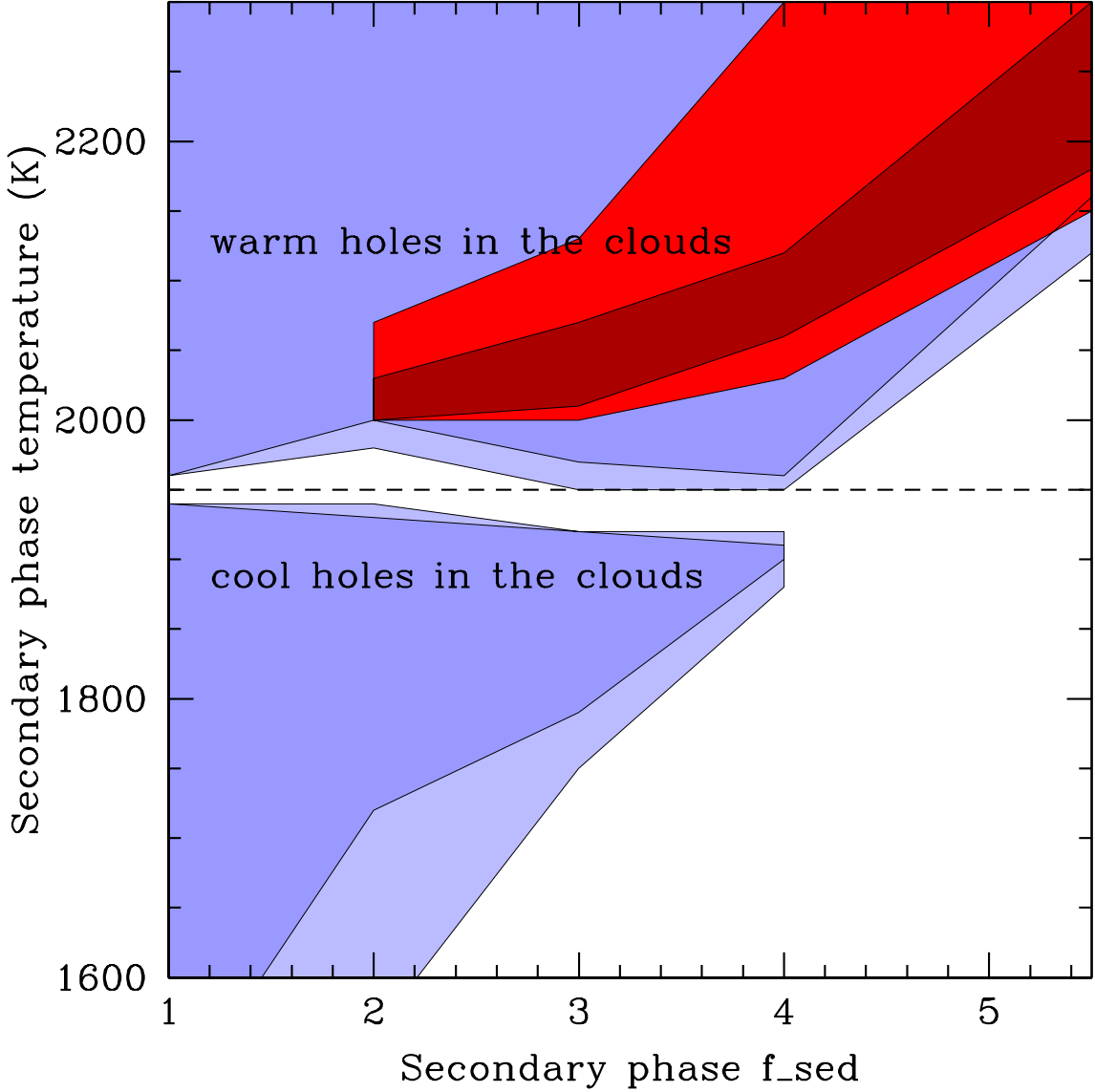}
\caption{Effective temperatures and cloud sedimentation parameter ($f_{\mathrm{sed}}$)
values permitted by our data for the secondary phase, provided the
primary phase has $f_{\mathrm{sed}} = 1$.  Cloudiness \textit{decreases} with
increasing $f_{\mathrm{sed}}$.  Blue-shaded regions are permitted
by the $J$/[3.6] amplitude ratio observed in our data, while red regions are
permitted by the [4.5]/[3.6] amplitude ratio, which turns out to be substantially
more constraining.  The darker shaded areas are
consistent with the data at the 1$\sigma$ level and the lighter areas at 2$\sigma$.
The $f_{\mathrm{sed}}$ values probed
are 1, 2, 3, 4, and $\infty$ (corresponding to a completely clear atmosphere).
As we cannot extend a plot axis to infinity, for purposes of illustrating
the models we have placed the $f_{\mathrm{sed}} = \infty$ results at 5.5 on
the $f_{\mathrm{sed}}$ axis.  The $T_{\rm eff} = 1950\,\rm K$ temperature of our
primary-phase model is indicated by the dashed horizontal line.
\label{fig:ampfig01}}
\end{figure}

\begin{figure}
\includegraphics[scale=0.8]{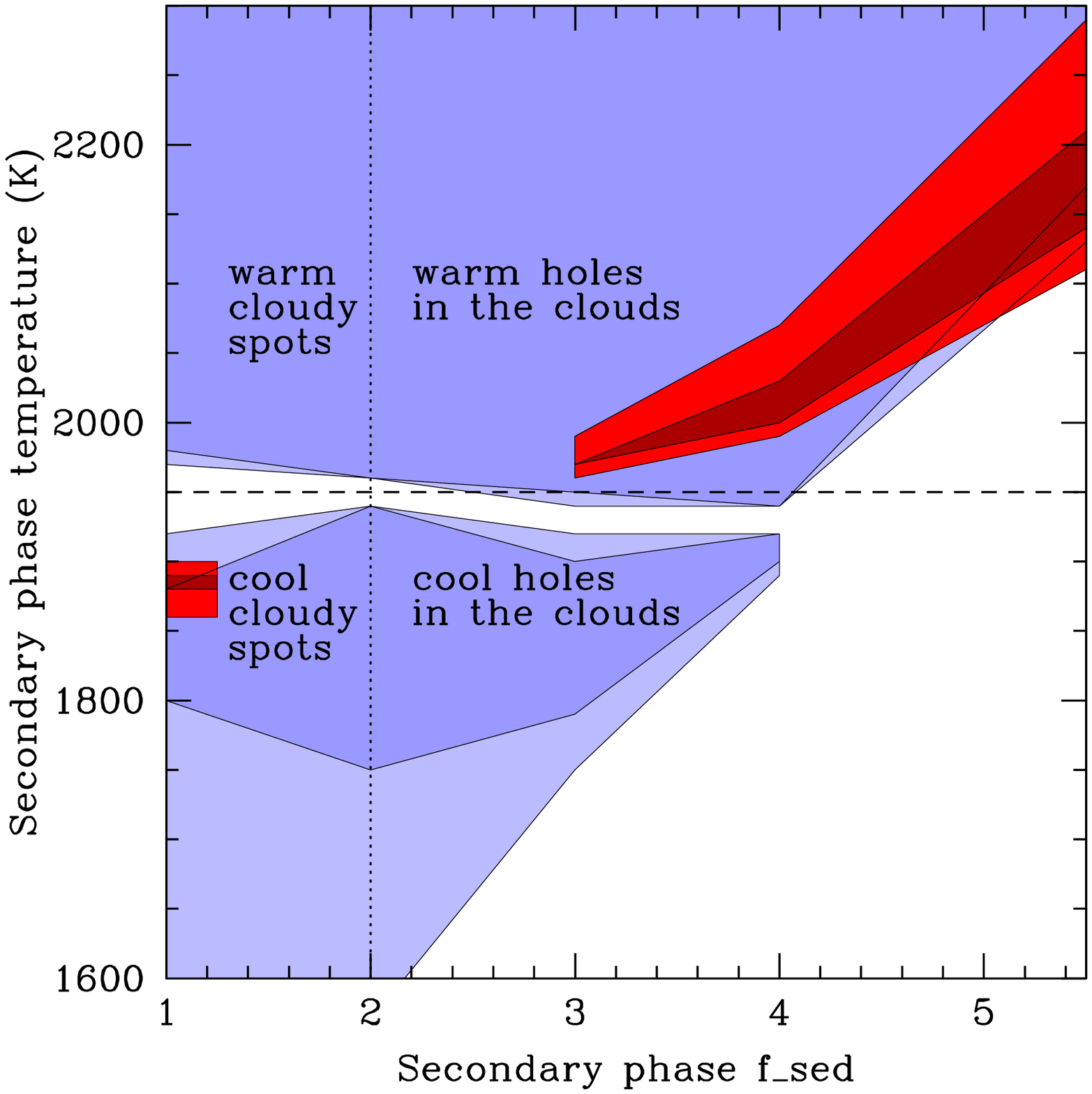}
\caption{Like Figure \ref{fig:ampfig01} but for a primary-phase $f_{\mathrm{sed}}$
value of 2.  The horizontal dashed line corresponds to the 1950 K $T_{\rm eff}$ of our
primary-phase model.  Points above it involve cloud anomalies warmer than the
primary phase, while points below it involve anomalies that are cooler.  The vertical
dotted line at $f_{\mathrm{sed}} = 2$ separates secondary-phase models with thicker
clouds than the primary phase (left of the line) from those with thinner clouds
(right of the line).\label{fig:ampfig02}}
\end{figure}

Our code calculates $\epsilon$ using Equation \ref{eq:amod02} (with straightforward adjustments
to account for the possibility that the hemisphere with only the primary phase
will actually be brighter), and then applies Equation \ref{eq:amod01}
to the other bands to predict the amplitude ratios.  For a given pair of
$f_{\mathrm{sed}}$ values for the primary and secondary phases, we seek values for
the temperature of the secondary phase which will simultaneously fit both
of our amplitude ratios.  We present our results in Figures \ref{fig:ampfig01}
and \ref{fig:ampfig02}, making a distinction between models that agree with
our data at the 1 and 2$\sigma$ levels.

Figure \ref{fig:ampfig01} shows the results if we take the primary phase
to have $f_{\mathrm{sed}} = 1$.  As $f_{\mathrm{sed}} = 1$ is the most heavily clouded model,
for this model the secondary phase must consist of a region where the cloud is less thick
or must involve a change in temperature only.  Secondary-phase temperatures can be found
that match all our data at the 1$\sigma$ level for $f_{\mathrm{sed}}$ values of 2, 3, 4, or
$\infty$.  The [4.5]/[3.6] amplitude ratio places stronger constraints
on the temperature than the $J$/[3.6] ratio; in fact, almost all models that match
[4.5] and [3.6] match our $J$-band results as well.  As noted above, there is no solution
if the secondary phase has the same value of $f_{\mathrm{sed}}$ as the primary
phase: cloud inhomogeneities are required to explain the data.
For this choice of the primary-phase model, the only solutions correspond to warm `holes' in the clouds:
that is, less cloudy patches in the global overcast that also have a higher $T_{\mathrm{eff}}$.  

Figure \ref{fig:ampfig02} shows the same analysis with a primary-phase
$f_{\mathrm{sed}}$ of 2.  Under this model, the secondary phase can have
either thicker or thinner clouds than the primary phase.  Warm hole-in-the-clouds
solutions exist as with our previous model.  However, here there
is also a small region of the parameter space, permitted at the 1$\sigma$ level, 
that corresponds to cold
regions of especially thick cloud: the signature we might expect if a cool, 
magnetic starspot has triggered increased condensate formation.  A model using
a primary-phase $f_{\mathrm{sed}}$ of 3 has the same broad
characteristics seen in Figure \ref{fig:ampfig02}:
solutions exist corresponding to either warm `holes' in the clouds or 
to cold regions of increased cloud thickness.

The full range in permitted values of $\epsilon$ (that is, the projected 
fraction of DENIS 1058's disk covered by the secondary phase) is 
0.8-11.0\% for solutions corresponding to warm `holes' in the clouds and 3-8\%
for solutions with cold regions of thicker cloud.  In both cases the smaller
values of $\epsilon$ correspond
to larger differences in $f_{\mathrm{sed}}$ between primary and secondary phases;
such scenarios also have the largest $T_{\mathrm{eff}}$ differences and produce
the highest brightness contrast.
For comparison, Voyager images show the Great Red Spot and its peripheral clouds 
covering about 3\% of Jupiter's visible disk.

It is worth noting that every scenario permitted by our data has the clearer 
phase at a higher $T_{\mathrm{eff}}$ than the cloudier phase.  This is true
whether the clearer phase is the secondary phase (localized, warm `holes' in the clouds)
or the primary phase (in which case the secondary phase consists of
localized cold regions of thicker cloud).  It makes
sense that we would see down to deeper, warmer layers of the atmosphere
in areas of reduced cloud opacity, and this
intuition is borne out by atmospheric physical considerations.
Well below any cloud decks the global atmospheric temperature everywhere on the 
brown dwarf must be essentially constant.   Compared to the nominal primary-phase
model ($T_{\rm eff} = 1950\,\rm K$, $f_{\mathrm{sed}} = 2$), a cloudless model with the same $T_{\mathrm{eff}}$
is roughly 250 K cooler at depth (pressure $P = 10\,\rm bars$).  Thus 
if the photosphere of a cloudy L dwarf were to relax to the thermal profile of a 
cloudless atmosphere with the same deep atmospheric thermal profile, we would 
expect that profile to be similar to a cloudless model with a $T_{\rm eff}$ about 
200 K warmer than the cloudy case and not similar to a cloudless model with the 
same $T_{\rm eff}$ as the cloudy case.  Thus the finding shown in Figure \ref{fig:ampfig02}, for 
example, that the cloudless model that pairs best with an $f_{\rm sed}=2$ model 
is 180 to 260 K warmer is fully consistent with this picture.  The findings 
that smaller and larger temperature differences are required for the cases of 
smaller and larger differences in $f_{\rm sed}$ (respectively) are likewise consistent with the 
atmospheric thermal profiles.  Nothing in our modeling method requires the
results to be consistent with this physical reasoning: they simply are.  The
same pattern has been consistently found in early T dwarfs:
\citet{SIMP0136}, \citet{2M2139}, and \citet{Apai2013} all found that their
data could be fit only if the temperatures of clearer regions were warmer
than those of cloudier regions.

The cloud inhomogeneities we observe could in principle be linked
to magnetic phenomena.  Magnetic heating in the atmosphere could evaporate
condensates and create warm `holes' in the clouds.  Similarly, cold
regions of thicker cloud could be `cloudy starspots' in which the formation
of increased condensates was triggered by a temperature reduction due to
the magnetic suppression of convective heat transport.  We note that
neither mechanism is necessarily required: Jupiter exhibits both cold
regions of thick cloud (e.g. the Great Red Spot, \citet{GelinoThesis})
and warm regions of unusually low cloud opacity, without requiring a
magnetic trigger for either.

Formally, the analysis in this section assumes that the $J$-band
variability of DENIS 1058 is due to the same set of cloud features as the variability
in IRAC ch1, and that the clouds did not change appreciably in the 
three days between the $J$-band and IRAC observations.  The time required for
substantial changes in the clouds of L dwarfs is an open question.
However, we note that the strongest constraints come from the IRAC
bands, and that the [4.5] data was taken immediately after that
at [3.6], rendering it less likely that changing cloud patterns could
have affected the measured amplitude ratio.

\subsection{Cold Magnetic Starspots} \label{sec:magspots}

Starspots form when locally strong magnetic fields inhibit convection.
This can happen only if the convecting gas is sufficiently ionized (i.e., electrically
conductive) to interact strongly with the field.
It is known based on radio observations that at least some L dwarfs 
have magnetic fields \citep{Lflares02,Lflares05,Berger09}, but
theory indicates that the cool, neutral atmospheres of even early L dwarfs are much too electrically
resistive for starspots to form \citep{nospots,Gelino02,Chabrier06}.  
In particular, \citet{Chabrier06} comment that the magnetic fields of L dwarfs should allow them to
have coronae (which may explain the observed radio emission) but not chromospheres.

Consistent with this theoretical picture,
indicators of magnetic activity such as H$\alpha$ emission and X-ray flux decline
as one goes from the mid M-stars down through the L dwarfs \citep{nospots,flare}.
We have already noted that DENIS 1058's H$\alpha$ emission does not
make it an exception to this general trend of decreasing activity in the L dwarfs:
its weak emission is not atypical for its spectral type.
Also interesting in this context is the fact that the well-known Benz-G\"{u}del relation 
connecting radio to X-ray flux in radio-emitting late type
stars is strongly violated for late M and cooler objects, such that for L dwarfs the
X-ray fluxes are $\sim 10^4$ times too faint relative to the radio \citep{Lflares05,Stelzer12}.
This may indicate that as temperature decreases, a profound change takes place in the way
the magnetic field interacts with the photosphere.  If, consistent
with theory, we attribute this to the atmosphere becoming uncoupled from
the magnetic field, it follows that L dwarfs cannot exhibit star spots.

However, the fact that H$\alpha$ emission, though very weak, does still exist
for DENIS 1058 (and some other L dwarfs) appears inconsistent with the theoretical
arguments that neither starspots nor chromospheres should be able to form.  
Several suggestions have been put forward to resolve this conundrum. \citet{nospots} have 
proposed that buoyant magnetic flux tubes could rise rapidly from ionized
regions deep in an L dwarf's interior and release their energy in the object's atmosphere,
which would produce H$\alpha$ emission in the absence of starspots.  Alternatively,
\citet{Lane07} attribute their detection of $I$-band variability in the radio-emitting
L3.5 dwarf 2MASS J00361617+1821104 to starspots, and get around the neutral-atmosphere
problem by proposing magnetic field intensifications across a large enough region
that when the field inhibits convection in deep, ionized layers of the star,
the effect is still seen at the photosphere.  Helling et al. (2011a,b)
propose that collisions between dust grains and/or lightning discharges in
brown dwarf atmospheres could produce enough ionization to couple the
atmosphere to the magnetic field.  Under this last scenario, it also seems plausible
that the ionization from lightning could be sufficient to explain the
observed H$\alpha$ emission without necessarily being enough to allow starspots.

This plethora of suggestions illustrates that while the interior magnetic dynamos
of fully convective objects such as brown dwarfs
and very low mass stars have been successfully modeled \citep{Chabrier06,Dobler06},
detailed models do not yet exist to constrain photospheric and
chromospheric magnetic phenomena in such objects.  In the absence of such models
we cannot definitively rule out magnetic starspots as an explanation for the
variability we have observed in DENIS 1058 --- although, as noted above,
cloud inhomogeneities are required in addition to starspots to explain our data.

\subsection{Magnetic Emission Regions} \label{sec:magsemm}

Magnetic fields can produce local emission regions in the form of aurorae, 
in which energetic electrons flow along external
magnetic field lines into an object's atmosphere.  Chromospheric emission
is produced by magnetic phenomena in denser gas closer to the photosphere,
but in some cases is similarly explained by energetic electrons impinging on
the gas (see for example the stellar flare model of Allred et al. 2006).

Although stellar flares have too short a characteristic 
timescale \citep{Lflares02, Mflares12} to account for 
the four-hour periodicity we observe in DENIS 1058, either aurorae
or lower-level chromospheric emission regions could in principle 
create rotationally modulated variability. Such variability in
magnetically-caused emissions
has been seen in L dwarfs. \citet{Berger09} saw periodic variations in both
radio and H$\alpha$ emissions from 2MASSW J0746425+200032.  The
H$\alpha$ equivalent width varied from 2.4 to 3.1 \AA,
which corresponds to a variation in the ratio of H$\alpha$ to
bolometric flux of roughly $6.3 \times 10^{-6}$ to $8.2 \times 10^{-6}$.
For comparison, H$\alpha$ emission from DENIS 1058 has been observed
at equivalent widths of $1.3\pm0.4$ \AA, 1.6 \AA, and $1.0 \pm 0.4$ \AA~ by 
\citet{noLi}, \citet{Kirk}, and \citet{martin99}, respectively.
Its H$\alpha$ flux is about $2.1 \times 10^{-6}$ of its bolometric flux \citep{flare}.


There are several difficulties with explaining our DENIS 1058
observations in terms of a magnetic emission region.  First, the only
known marker of magnetic activity in DENIS 1058, the H$\alpha$ emission,
shows no evidence for variability.  Although
only large H$\alpha$ variations ($\gtrsim 50$\%) would have been detected,
much larger variations in H$\alpha$ than in broadband flux
are to be expected simply because all of the
H$\alpha$ emission is magnetically generated while most of the IR continuum is not.
Second, even supposing the H$\alpha$ emission to be variable at an
undetectable level, the flux variations we
observe may be too large relative to the
measured H$\alpha$ flux to be reasonably explained by the same emission regions.
Third, granting a scenario in which a very large [3.6]/H$\alpha$ 
emission ratio is possible, it remains difficult to explain the $J$-band variation by the same phenomenon.
We expand on these latter two points below.

The IRAC [3.6] magnitude that we measure for DENIS 1058 corresponds
to a flux of about $9.9 \times 10^{-16}$ W m$^{-2}$, and the 2MASS $J$ magnitude
of 14.16 corresponds to a flux of $1.1 \times 10^{-15}$ W m$^{-2}$.  These fluxes
comprise 9\% and 10\%, respectively, of the $1.1 \times 10^{-14}$ W m$^{-2}$ bolometric
flux found by \citet{spec01}.  The 0.388\% and 0.843\% variations that we
observe in these wavelengths therefore correspond to $3.5\times10^{-4}$ and $8.5\times10^{-4}$
of the bolometric flux, respectively --- 170 and 400 times larger than the total
observed H$\alpha$ flux.  If a localized magnetic emission region is
responsible for the variability we observe, its excess luminosity in
both IRAC [3.6] and the $J$ band must be more than two orders of magnitude greater
than the entire H$\alpha$ emission from DENIS 1058.  

As regards the IRAC [3.6] band, such a luminosity ratio is not 
necessarily inconsistent with an aurora.  
Aurorae on the giant planets of our own solar system produce
line emission at Lyman-$\alpha$ \citep{JHA02,LyAlph01}, 
H$\alpha$ \citep{JHA01,SHA01}, and in the near infrared emission lines
of the $\mathrm{H_3^+}$ molecule.  The latter is detectable at 
2$\mu$m wavelengths \citep{H3a}, but stronger in its fundamental band
around 4$\mu$m \citep{H3b,H3c}, where its luminosity
can indeed be more than two orders of magnitude greater than the entire optical
(i.e. H$\alpha$) auroral luminosity \citep{aurora_review}.

However, no $J$-band emission lines are observed in the Jovian aurora.  Thus, while Jovian-like
aurorae on DENIS 1058 might explain its H$\alpha$ emission and its [3.6] variability,
emission lines not seen in such aurorae would be required to explain
the even larger variability we see in the $J$-band.  Different molecules and different emission
lines could exist in the much warmer DENIS 1058, but there are also
observational constraints on magnetically-caused near-IR emission in warmer objects.  
\citet{Stelzer12} obtained simultaneous optical and 
near-IR spectroscopy of the active M9 dwarf DENIS-P J104814.7-395606 at a time
when it was showing emission in the hydrogen Balmer lines out to $n=8$, as well as in 
the Ca II H and K lines.
The circumstances were ideal for detecting emission lines in the $J$ band if 
any existed, but emission lines were absent not only in the $J$ band but
throughout the near-IR, including the 2$\mu$m regime where $\mathrm{H_3^+}$
lines are seen in Jupiter. These results for objects bracketing DENIS 1058 in temperature suggest
that the $J$-band variability we observe cannot readily be explained
in terms of line emission from a magnetically heated region.

None of the difficulties we have outlined above is sufficient to 
conclusively rule out auroral or chromospheric emission as the cause
of DENIS 1058's variability.  Although we have shown in
\S \ref{sec:mod} that blackbody continuum emission from
a magnetically heated region cannot explain the amplitude ratios we observe,
some combination of blackbody and line emission could conceivably do so.
It is also possible that an auroral electron beam or other release of magnetic
energy could evaporate condensates and create a clearing in the clouds: thus our observations
could be due to interactions between magnetic and cloud phenomena.
A detailed theory of UCD aurorae, as well as additional observations,
is required to constrain such possibilities.  At present, we note that
a fully self-consistent explanation involving inhomogenous clouds is
possible and seems to involve fewer difficulties than scenarios involving only magnetic emission.

\section{Conclusion} \label{sec:conc}
DENIS 1058 exhibits periodic photometric variability in IRAC [3.6] with
a period of $4.25^{+0.26}_{-0.16}$ hours and a peak-to-valley amplitude of $0.388 \pm 0.043$\%.
In the $J$-band we measure variations with a larger peak-to-valley
amplitude of $0.843 \pm 0.098$\%, and a period of $4.31 \pm 0.31$ hours, which is
consistent with our [3.6] results at the 1$\sigma$ level.  DENIS 1058 may
exhibit very weak IRAC [4.5] variability positively correlated with that at [3.6],
but our measured [4.5]/[3.6] amplitude ratio of $0.23 \pm 0.15$ is
consistent with zero variability at [4.5].

The different amplitudes we detect at different wavelengths rule out pulsation or
tidal distortion due to a close binary as the cause of DENIS 1058's periodic
variations; a pulsational interpretation is further ruled out because the
period is much too long.  This implies that the variability
we detect is rotationally modulated.  Our photometric
period may thus be combined with published $v \sin(i)$ results to
obtain a 2$\sigma$ lower limit of $0.111 ~R_{\sun}$ on DENIS 1058's radius.
This implies a mass and age less than $0.055 ~M_{\sun}$ and 320 Myr, respectively,
values which are consistent with the young-object kinematics
noted by \citet{martin97} and \citet{Dahn02}, but not with the lithium non-detections
of \citet{martin97}, \citet{noLi}, and \citet{Kirk}, nor with the luminosity measurements of
\citet{spec01} and \citet{Dahn02}.  The age, mass, and radius limits could
be reconciled with the luminosity if the effective temperature were $2.7 \sigma$
cooler than measured, though the absence of lithium would remain
puzzling.  Alternatively, the radius limit would change to a value consistent
with the observed luminosity and $\gtrsim 800$ Myr age implied by the
lithium results if the true rotational velocity were $3.3 \sigma$ less than
has been measured.  Regardless of which (if either) of these scenarios
applies, the large radius limit indicates a near-equatorial viewing
geometry: i.e., DENIS 1058's rotation axis is probably inclined
substantially more than $45^{\circ}$ to our line of sight.

We have modeled DENIS 1058's variability under the assumption that it
is due to cloud inhomogeneities and/or photospheric starspots.  
Our two-phase cloud models yield viable solutions in which the inhomogeneities
take the form of warm `holes' in a global overcast: that is, regions where the clouds are thinner and the
$T_{\mathrm{eff}}$ is higher.  A smaller number of solutions exist involving regions of
even thicker, cooler cloud within the global overcast: the scenario we would
expect if cold magnetic starspots have triggered the cloud formation.  
In the absence of cloud inhomogeneities, neither cold nor warm magnetic
spots are able to fit our observations.
We note that there is not yet a consensus on whether magnetic starpots can occur
in L dwarfs.

For every model that can fit our data, the $T_{\mathrm{eff}}$
of the clearer phase is warmer than that of the cloudier phase.
Our data cannot be explained by clearings that are colder than
the surrounding clouds nor by thicker cloud patches that are warmer than their surroundings.
The same result has been found in similar analyses of early T dwarfs 
by \citet{2M2139}, \citet{SIMP0136}, and \citet{Apai2013}.  
This is consistent with physical considerations regarding the model atmospheres:
Deep in a brown dwarf's interior, the pressure and temperature under both clearer
and cloudier regions must be the same, and if we compare clearer and
cloudier model atmospheres with the same deep adiabat,
we find that the clearer models invariably have a higher $T_{\mathrm{eff}}$.

Magnetic emission regions in L dwarfs can create variability at radio
wavelengths and in the H$\alpha$ emission line, and in principle could
also cause variations at the wavelengths we have observed.  Difficulties
with such an interpretation include the fact that H$\alpha$ observations
of DENIS 1058 have shown no evidence of variability; that the emission
would have to be very efficient in IRAC [3.6] and the $J$-band relative to H$\alpha$;
and that no emission lines capable of explaining the $J$-band variations
we observe are readily apparent in spectra of other objects with either auroral
or chromospheric emission.  None of these difficulties is necessarily fatal,
and further observations and theoretical modeling are required to understand
L dwarf aurorae.  At present, however, explaining the variations of DENIS 1058 by
inhomogenous clouds (which might be coupled to magnetic phenomena) appears
to involve the fewest difficulties.

\section{Acknowledgements} 
We thank Didier Saumon for supplying us with files containing the model spectra
of \citet{marley}, which we have used to construct our two-phase models
of DENIS 1058.  This research was 
supported by NASA through the Spitzer Exploration Science Program {\it 
Weather on Other Worlds} (program GO 80179) and ADAP award NNX11AB18G.
This research is also based on observations obtained at the 
Southern Astrophysical Research (SOAR) telescope, 
which is a joint project of the Minist\'{e}rio da Ci\^{e}ncia, Tecnologia, 
e Inova\c{c}\~{a}o (MCTI) da Rep\'{u}blica Federativa do Brasil, 
the U.S. National Optical Astronomy Observatory (NOAO), the 
University of North Carolina at Chapel Hill (UNC), and Michigan State University (MSU).
The SOAR observations reported herein were made under Chilean program CN2012A-055. 
Radostin Kurtev acknowledges support from Proyecto DIUV23/2009,
Centro de Astrof\'{i}sica de Valpara\'{i}so, and FONDECYT through grant 1130140.
We thank Nikole Lewis for supplying us with IDL code
to measure the noise pixel parameter of our data, for
allowing us to read the draft version of her paper
describing uses of this parameter in IRAC photometry,
and for additional helpful advice.
This publication makes use of the SIMBAD online database,
operated at CDS, Strasbourg, France, and the VizieR online database (see \citet{vizier}).
This publication makes use of data products from the Two Micron All Sky Survey, 
which is a joint project of the University of Massachusetts and the Infrared Processing
and Analysis Center/California Institute of Technology, funded by the National Aeronautics
and Space Administration and the National Science Foundation.
We have also made extensive use of information and code from \citet{nrc}. 
We have used digitized images from the Palomar Sky Survey 
(available from \url{http://stdatu.stsci.edu/cgi-bin/dss\_form}),
 which were produced at the Space 
Telescope Science Institute under U.S. Government grant NAG W-2166. 
The images of these surveys are based on photographic data obtained 
using the Oschin Schmidt Telescope on Palomar Mountain and the UK Schmidt Telescope.

Facilities: \facility{SOAR, Spitzer}

\section{Appendix A: Iterative Procedure for Improving Relative Photometry}
We begin with ordinary relative photometry of DENIS 1058 and the reference
stars, constructed using Equation \ref{eq:rp01} with Stars 7 and 9 dropped
from consideration.  We normalize the resulting photometry.  We then seek to
fit a function by which the raw photometry of a given reference star may be divided to remove systematic
errors specific to that star and leave only the effects common to all stars.
We have already determined that, aside from the nod-position
offset, no physically motivated model of systematic errors does as well as
a linear function of time.  As we are now trying to remove all systematic effects, 
a function more complex than a linear trend is warranted.  We
conservatively choose only a quadratic in order to ensure that it cannot
mimic a sinusoid by producing two local extrema internal to the data sequence.
We seek to model
the normalized relative photometry of each reference star using a least-squares 
SVD fit to a function of time and nod position having the form:

\begin{equation}
f_k(t) = B_{nod} + A_0 + A_1t + A_2t^2
\tag{A1} \label{eq:rp02}
\end{equation}

where $B_{nod}$ is zero for images taken in nod position 1 and has a constant
value for images taken in nod position 2, and all three parameters will,
of course, have unique values for each reference star.  We identify outliers
more than 2.5$\sigma$ from the fit as bad points.  No more than 5 points are
rejected this way in any iteration on any star.  Over good points only, we
create an adjusted version of the raw photometry for this reference star:

\begin{equation}
G_{ki} =  \frac{F_{ki}}{f_k(t)}
\tag{A2} \label{eq:rp03}
\end{equation}

Note that because $f_k(t)$ was obtained through a fit to normalized photometry, it
is never far from 1.0 for any value of $t$, and therefore $G_{ki}$ will differ
only subtly from $F_{ki}$.  It should, however, differ in the sense that the
systematic errors specific to star $k$ will have been substantially reduced.
It remains to get values for $G_{ki}$ for points corresponding to the bad
points.  Over good points only, we construct relative photometry as follows:

\begin{equation}
S_{ki} =  \frac{G_{ki}}{\sum_{j=2, j \ne k}^{m} F_{ji}}
\tag{A3} \label{eq:rp04}
\end{equation}

Then we replace bad points using:

\begin{equation}
G_{kb} =  <S_{ki}> \sum_{j=2, j \ne k}^{m} F_{j,b}
\tag{A4} \label{eq:rp05}
\end{equation}

where $b$ indicates a specific image on which the photometry of star $k$
was bad, and $<S_{ki}>$ is an average over all $i$ where the photometry
of star $k$ was good.  Effectively, the summed fluxes of all non-deviant
stars on image $b$ are being used to create a proxy value to replace the 
deviant photometry of star $k$ on this image.  When the photometry of all
reference stars has been corrected using Equations \ref{eq:rp03} and \ref{eq:rp05},
we construct a new, second-iteration version of the relative photometry $R_{ki}$: 

\begin{equation}
R_{ki} =  \frac{F_{ki}}{\sum_{j=2, j \ne k}^{m} G_{ji}}
\tag{A5} \label{eq:rp06}
\end{equation}

We proceed with the second iteration,
fitting this new relative photometry (after normalization) using Equation \ref{eq:rp02}.
The second and subsequent iterations are the same as the first except that in Equations \ref{eq:rp04}
and \ref{eq:rp05}, we can now use the adjusted raw photometry $G_{ji}$ where 
$F_{ji}$ appeared before.  Note, however, that the original raw photometry is always
used in the numerator of Equation \ref{eq:rp06} to construct the new relative photometry
at the start of each iteration.  Thus in each iteration, the fit to Equation \ref{eq:rp02}, and the
identification of outliers from this fit, proceeds independently of the fits or outliers found
in previous iterations.  Points that were
found deviant on one iteration may (if the deviance was not intrinsic but was due to as
yet uncorrected bad photometry in another star) be accepted on a subsequent iteration,
and multiple `layers' of quadratic fits cannot add up to yield, effectively, a fit of
much higher order.

\end{document}